\documentclass[pre,aps,superscriptaddress]{revtex4}
\usepackage{graphicx}

\begin{document}

\title{Discrete solitons and vortices
in
hexagonal and honeycomb
lattices: \\ existence, stability and dynamics}
\author{K.J.H. Law}
\affiliation{Department of Mathematics and Statistics, University of
Massachusetts, Amherst MA 01003-4515}
\author{P.G. Kevrekidis}
\affiliation{Department of Mathematics and Statistics, University of
Massachusetts, Amherst MA 01003-4515}
\author{V. Koukouloyannis}
\affiliation{School of Physics, Theoretical Mechanics, Aristotle
University of Thessaloniki,54124 Thessaloniki Greece}
\affiliation{Department of Civil Engineering, Technological Educational Institute of Serres, 64124 Serres, Greece}
\author{I. Kourakis}
\affiliation{Centre for Plasma Physics, Queen's University
Belfast,  BT7 1 NN Northern Ireland, UK}
\author{D.J. Frantzeskakis}
\affiliation{Department of Physics, University of Athens,
Panepistimiopolis, Zografos, Athens 15784, Greece}
\author{A.R. Bishop}
\affiliation{Theoretical Division and Center for Nonlinear Studies,
Los Alamos National Laboratory, Los Alamos, New Mexico 87545, USA}
\begin{abstract}
We consider a prototypical
dynamical lattice model, namely
the discrete nonlinear Schr{\"o}dinger equation on non-square lattice
geometries. We present a systematic classification of the solutions
that arise in principal six-lattice-site and three-lattice-site contours
in the form of both discrete multi-pole solitons
and discrete vortices.
Additionally to identifying the possible states,
we analytically track their linear stability both qualitatively and
quantitatively. We find that among the six-site configurations, the
``hexapole'' of alternating phases ($0$-$\pi$), as well as the vortex
of topological charge $S=2$ have intervals of stability; among three-site
states, only the vortex of topological charge $S=1$ may be stable in
the case of focusing nonlinearity. These
conclusions are confirmed both
for hexagonal and for honeycomb lattices by means of detailed numerical
bifurcation analysis of the stationary states from the anti-continuum
limit, and by direct simulations to monitor the dynamical instabilities,
when the latter arise. The dynamics reveal a wealth of nonlinear
behavior resulting not only in single site solitary waveforms, but
also in robust multi-site breathing structures.
\end{abstract}

\maketitle

\section{Introduction}
\label{intro}

Hamiltonian lattice or quasi-discrete systems have become
popular in the last few years, to a considerable extent due to experimental
implementations of such systems
drawn from various
branches of physics.
One of the first examples where such developments became relevant
was in the nonlinear optics of fabricated AlGaAs
waveguide arrays \cite{7}. There, the interplay of inherent
discreteness and nonlinearity led to the emergence of numerous
interesting phenomena including
Peierls-Nabarro potential barriers, diffraction
and diffraction management \cite{7a},
gap
solitons \cite{7b}, and so on (see also the reviews \cite{review_opt,general_review}
and references therein).

On the other hand, more recently another area of nonlinear optics
that has been central to the development of both theoretical as well
as of computational tools to study such systems, has been the
setting of optically induced photonic lattices in
photorefractive crystals such as SBN. There, the original theoretical
proposal of
these lattices \cite{efrem} was
susbequently followed by experimental realizations
\cite{moti1,moti2}, paving the way for the observation of a
diverse array of novel and interesting phenomena in such crystals.
These include the formation of patterns such as dipole \cite{dip},
quadrupole \cite{quad} and necklace \cite{neck} solitons, impurity modes
\cite{fedele}, discrete vortices \cite{vortex1,vortex2}, rotary solitons
\cite{rings}, higher order Bloch modes \cite{neshev2} and gap vortices \cite{motihigher},
the observation of two-dimensional (2D)
Bloch oscillations and Landau-Zener tunneling \cite{zener},
the observation of localization and diffraction in honeycomb \cite{honey},
hexagonal \cite{rosberg2}
and quasi-crystalline
lattices \cite{motinature1}, and most recently the study of
Anderson localization in disordered photonic lattices \cite{motinature2}
(for a review of
some of this activity see, e.g., Refs.
\cite{moti3,zc}).

Finally, we should note
that similar dynamical lattices have also become
of interest in an entirely different area of physics, namely the
atomic physics of Bose-Einstein condensates (BECs), when trapped in periodic
potentials; see, e.g., the recent reviews \cite{konotop,markus2,ourbook}.

In this work, we will focus, in particular, on the recently
emerging area of {\it non-square} lattices in
waveguide arrays, as well as in light induced photonic crystals,
\cite{honey,rosberg2,szameit,tja2,bam}.
Such lattices were also considered earlier from a theoretical perspective
in discrete settings corresponding to waveguide arrays
\cite{gaid}.
We will study here
the existence, stability and
dynamical properties of multi-pulse solitary wave structures, as well as of
discrete vortex structures in both hexagonal and honeycomb lattices.
We will focus on two prototypical contours of such lattices.
Namely, a more extended six-site contour, as well as a reduced
three-site contour, both
depicted in Fig.  \ref{lattice}.
Our prototypical model of interest will be the discrete
nonlinear Schr{\"o}dinger (DNLS) equation and our results
will be presented for the case of a focusing nonlinearity;
however, our findings can be directly transformed
to the case of a defocusing nonlinearity. Additionally, we should
note that similar results can be obtained in Klein-Gordon models
and have been illustrated,
e.g., for three-site contours in hexagonal lattices \cite{kouk}.

It is relevant to note
that crystalline configurations of strongly coupled doped plasmas
(dusty plasma crystals) occur in the form of 1D or 2D monolayers formed
in low-temperature gas discharge experiments \cite{IKref1}.
Interestingly, such dust crystals generically appear as spontaneously formed
hexagonal 2D arrangements \cite{IKref2}, although alternative configurations
also  include honeycomb 2D lattices \cite{IKref3} and 1D dust
chains (when appropriate
trapping potentials are used for lateral confinement \cite{IKref4}).
A discrete Klein-Gordon description has recently been employed
to model the dynamics
of transverse vibrations of dust grains in dusty plasma crystals,
both in 1D \cite{IKref5} and in hexagonal 2D dust lattices \cite{IKref6}.

The key findings that we report here are the following:
\begin{itemize}
\item For the focusing nonlinearities considered herein, in a six-site
honeycomb/hexagonal contour, topological charge $S=2$ configurations
may be stable, while $S=1$ ones can never be stable. This represents
a notable qualitative difference from the results in the case of a square
lattice \cite{peli2d}, where the prototypical contour consisting
of four sites features a potentially stable $S=1$ vortex, as well
as a genuine
potentially stable analog of the $S=2$ vortex
(the so-called $S=2$ quasi-vortex
\cite{kchen}).
\item In these contours, also six-site alternating $0$-$\pi$ phase
configurations are potentially stable, while in-phase configurations are
not stable. While this instability can be implicitly inferred from
the instability of the corresponding building blocks (i.e., the
instability
of the in-phase dipole and the potential stability of the out-of-phase
dipole \cite{jpa_bam}), its quantitative characteristics can only be
traced through the approach presented below.
\item In three-site contours, the only potentially stable configuration
is that of a discrete vortex, while both in-phase and alternating phase
configurations are observed to be unstable.
\item The
evolution of the dynamical instability
in these lattices is more complex than in the square lattice case,
and may involve not
only degeneration
to single-site solitons but possibly to multi-site solitary wave structures,
and often the formation of robust breathing states, consisting of
multiple sites (possibly even as many as the original configuration).
In fact two clear breather formations recur in multiple simulations:
\begin{itemize}
\item Two sites with fluctuating, usually opposite, relative phases and
oscillating amplitudes of comparable magnitude.
\item Two sites with different amplitudes oscillating between the same
relative phases and opposite relative phases, depending on whether the
amplitudes are closer or further, respectively.
\end{itemize}
\end{itemize}
%
Six-site configurations with phases of $0$ or $\pi$ will be collectively
called ``hexapoles'' herein, while three-site configurations with phases
$0$ or $\pi$ will be collectively
termed ``tripoles''.

Our presentation of the above
findings is structured as follows. In section II, we
provide the background for the theoretical analysis. In section III,
we corroborate the theoretical findings with numerical bifurcation results
illustrating the various nonlinear modes in both hexagonal
and honeycomb geometries and their stability properties, while the
corresponding
dynamics are presented in section IV.
Finally, in section V, we summarize our
findings and present our conclusions.

\section{Theoretical Analysis}
\label{theory}

We consider the
following discrete nonlinear Schr{\"o}dinger equation (DNLS)
for the two geometries of interest:
\begin{eqnarray}
i \frac{du_{m,n}}{dz}= -\varepsilon \Delta_d - |u_{m,n}|^2 u_{m,n},\\ \label{dnls_eq1}
\Delta_d=\left(\sum_{\langle m',n'\rangle\in N} u_{m',n'} - |N|
u_{m,n}\right),
\end{eqnarray}
where the summation is over the set $N$ of nearest neighbors
(denoted by $\langle m',n'\rangle$) of the site $(m,n)$, $|N|$
is the cardinality
of the set $N$ of neighboring sites (six in the hexagonal geometry
and three in the honeycomb), and $u_{m,n}$ models, e.g.,
the envelope of the electric field in the corresponding waveguide
\cite{gaid}, while $\varepsilon$ represents the coupling strength
between nearest neighbor nodes; $z$ denotes the propagation distance
along the crystal.

\begin{figure}[h!]
\begin{center}
\includegraphics[width=40mm]{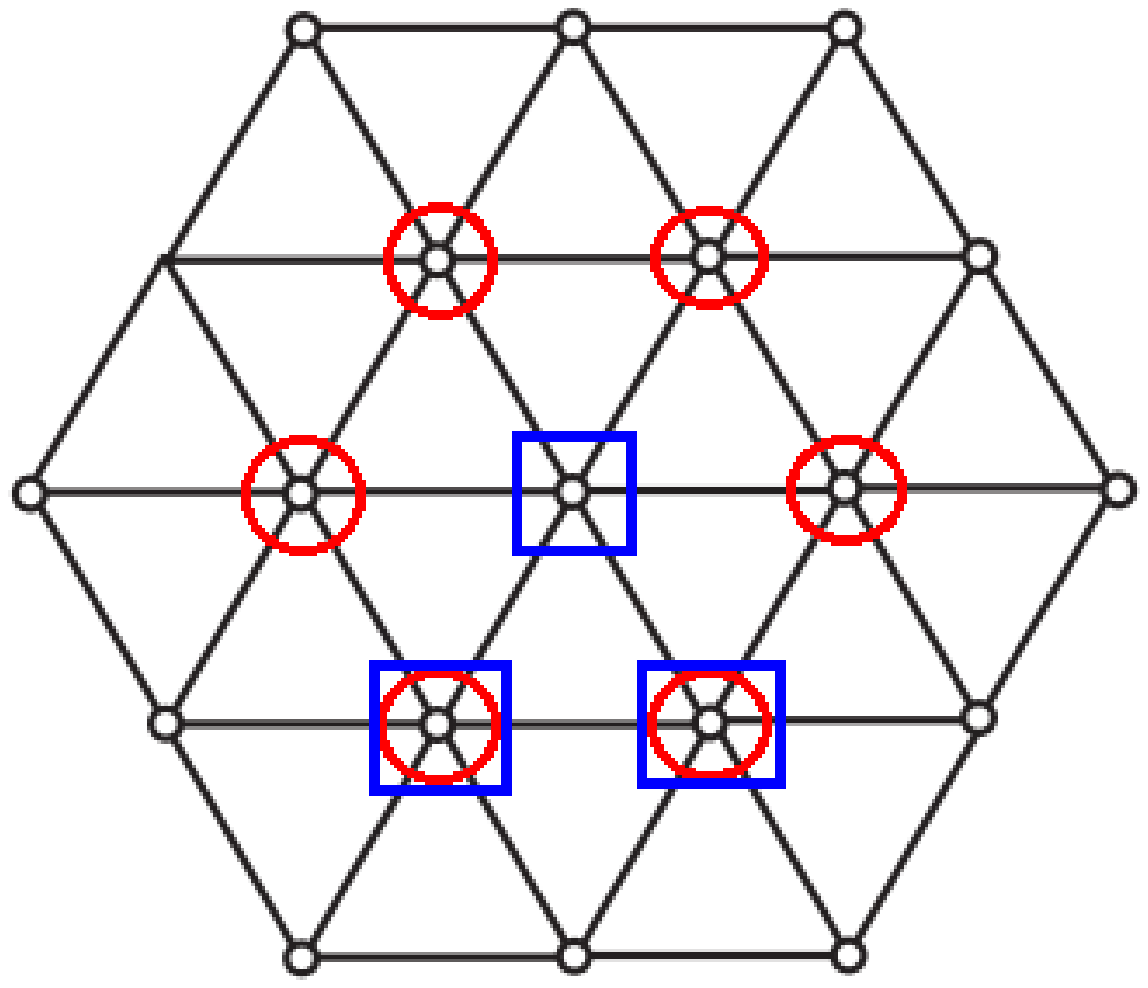}
\hskip 10mm
\includegraphics[width=40mm]{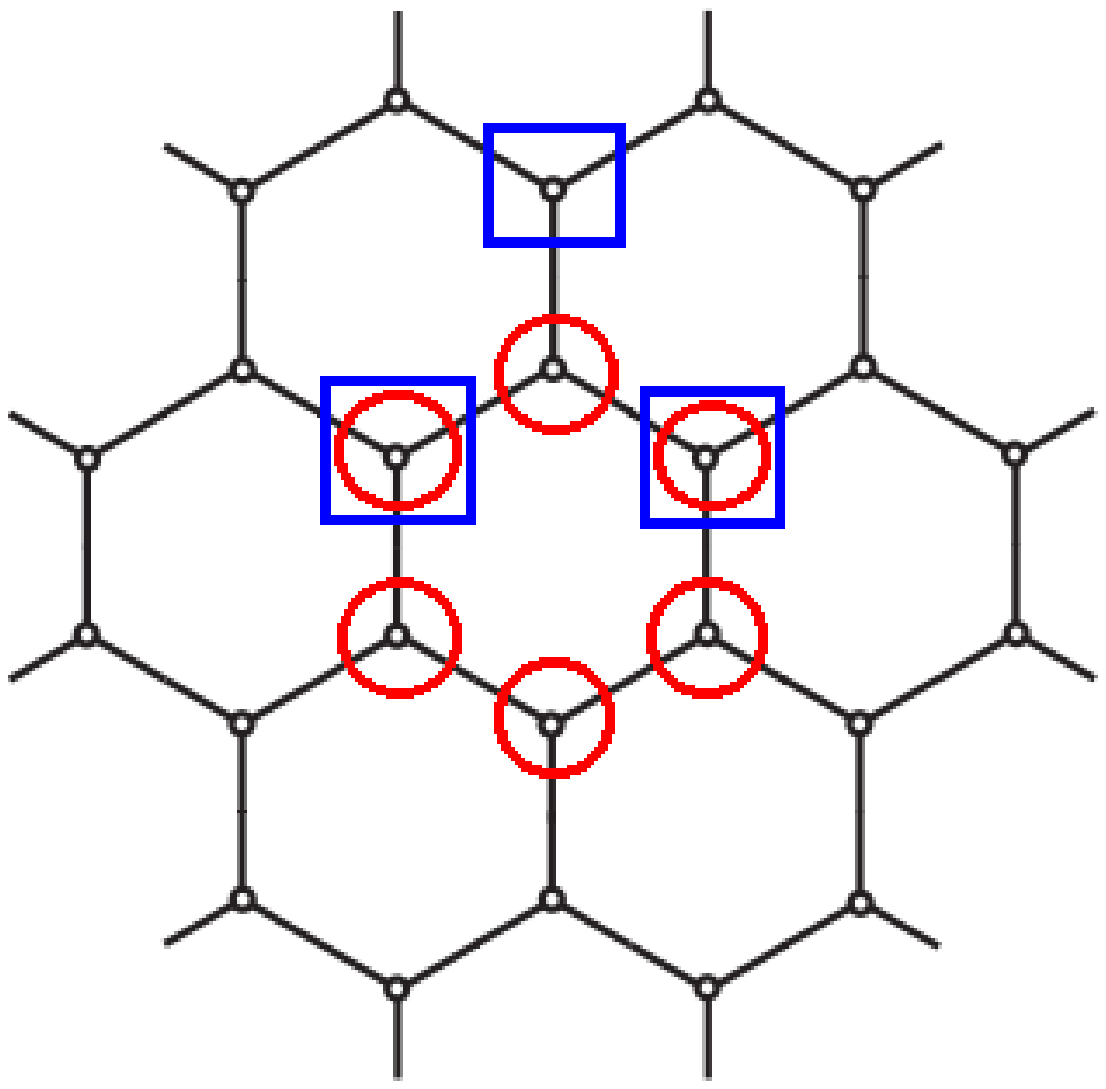}
\end{center}
\caption{(Color online)
Discrete lattice configurations for the hexagonal geometry (left),
in which each node has six neighbors, and the honeycomb geometry (right),
in which each node has three neighbors. The relevant ``hexapole''
configurations are represented by the red circles and
the ``tripoles'' are given by blue squares.
Notice that the relevant three site configuration
for the honeycomb is composed of next-nearest neighboring sites.}
\label{lattice}
\end{figure}

In the, so-called, anti-continuum (AC) limit
$\varepsilon \rightarrow 0$ the sites are uncoupled.
In this case, explicit solutions over contours of
nodes indexed by $j$ can be easily
found in the general form
$u_j = \sqrt{\Lambda}\exp(i\theta_j)\exp(i\Lambda z)$,
where $\Lambda$ is the propagation constant and $\theta_j \in [0,2\pi)$.
Without loss of generality we will fix the value of $\Lambda$, taking
$\Lambda=1$,
and we will consider three- and six-site contours in each of the hexagonal
and honeycomb geometries, shown in Fig. \ref{lattice}. According to the earlier
work of \cite{peli2d,peli1d} (see also \cite{suko}), the necessary condition
for a discrete contour $M$ in the AC limit to persist in the presence
of non-zero coupling is:
\begin{eqnarray}
g_j\equiv \sin(\theta_j-\theta_{j+1}) + \sin(\theta_j-\theta_{j-1}) =0,
\label{condition}
\end{eqnarray}
for all $j \in M$. The stability can also be determined from the
eigenvalues $\gamma_j$ of the $|M| \times |M|$
{\it Jacobian} ${\cal J}_{jk}=\partial g_j/\partial \theta_k$.  In
particular, for each eigenvalue $\gamma_j$, the full linearization of
Eq. (\ref{dnls_eq1}) around a stationary solution with non-zero nodes
in $M$ will have eigenvalue pairs $\lambda_j$ given, to leading order, by
the following relation in the case that the sites in $M$ are nearest
neighbors:
\begin{eqnarray}
\lambda_j = \pm \sqrt{2 \gamma_j \varepsilon}.
\label{eigs_eq3}
\end{eqnarray}

If the non-zero sites comprising the contour are next-nearest neighbors
instead, as in the case of the three site contours for the honeycomb
lattice geometry (see Fig. \ref{lattice}), then $\varepsilon$ is replaced
by $\varepsilon^2$ in the previous relation. Unstable solutions for weak
coupling (small $\varepsilon$) can then be
identified as those for which the eigenvalues $\lambda_j$ have
non-zero real part, given the Hamiltonian nature of the model.
The Jacobian matrix has the following form:
\begin{eqnarray}
\label{Melements3}
\begin{array}{lcl}
({\cal J})_{j,k} = &
 \left\{ \begin{array}{lcl}
\cos(\theta_{j+1} - \theta_j) + \cos(\theta_{j-1} - \theta_j), &
\quad & j = k, \\ - \cos(\theta_j - \theta_k), & \quad & j = k \pm 1, \\
0, & \quad & |k - j | \geq 2. \end{array} \right.
\end{array}
\end{eqnarray}

We will consider primarily contours $M$ such that
$|\theta_{j+1}-\theta_j|=\Delta \theta$
is constant for all $j \in M$,
$|\theta_1-\theta_{|M|}|=\Delta \theta$ and
$\Delta \theta |M| = 0$ ${\rm mod}$ $2\pi$,
except one case which will be treated separately.
In the primary case, all the non-zero elements of this
matrix are then factors of
$a=\cos(\Delta \theta)$, and the eigenvalue problem of the Jacobian
${\cal J}$ reduces to the following difference equations:
\begin{eqnarray}
a (2 x_n- x_{n+1} -x_{n-1})=\gamma_j x_n.
\label{diff_eq4}
\end{eqnarray}
These can be solved by a discrete Fourier transform with any eigenvector
$x_n \sim \exp(i 2\pi n j/|M|)$, whence $\gamma_j=4 a \sin^2(\pi j / |M|)$
and then
\begin{eqnarray}
\lambda_j=\pm \sqrt{8 \varepsilon
\cos\left(\Delta \theta\right) \sin^2\left(\frac{\pi j}{|M|}\right)}.
\label{stab}
\end{eqnarray}
Recall that 
for the honeycomb three-site next-nearest-neighbor
contours the above formula should be used
with $\varepsilon$ replaced by $\varepsilon^2$.
The special case of the three-node contour with phases
$0$, $\pi$ and $0$ can be treated also in the framework of the Jacobian
of Eq. (\ref{Melements3}) [and its eigenvalues computed by Eq.
(\ref{eigs_eq3})], although it does not fall under the general calculation
of Eqs. (\ref{diff_eq4})-(\ref{stab}).
We will consider nodes separated by either
$\Delta \theta=0,\pi, \pi/3$, or $2 \pi/3$,
for the different contours in this work.

These predictions for the linear stability eigenvalues will be compared
to numerical results for the {linear} stability of the stationary solution
$v_{m,n}\exp(iz)$ of Eq. (\ref{dnls_eq1}). The stability will be analyzed
by using the following ansatz,
%
\begin{eqnarray}
\begin{array}{lcl}u_{m,n}=
e^{iz}
 \left[v_{m,n} + \delta \left( p_{m,n}
e^{\lambda z} + q_{m,n}^{\star} e^{\lambda^{\star} z} \right) \right],
\label{linearization}
\end{array}
\end{eqnarray}
and solving the ensuing eigenvalue problem for the eigenvalue
$\lambda$ and the eigenvector $[p_{m,n},q_{m,n}]^{T}$.
In the above, the asterisk ($\star$) denotes complex conjugate and
``$T$'' denotes transpose. Notice that when the resulting
eigenvalues of the linearization possess a nonzero real
part, the solution will be exponentially unstable, with a growth
rate corresponding to the real part of the most unstable eigenvalue.

\section{Numerical results
}
\label{ex_stab}

Both in this section, detailing the various configurations and
their corresponding stability over the six-node and three-node
contours, and in the next one, comparing the corresponding
dynamics, we will partition our discussion to two subsections.
The first one will be devoted to the results obtained for
the hexagonal geometry, and
the second
devoted to the case of the honeycomb geometry.

\subsection{Hexagonal geometry}
\label{hex_exstab}

\begin{figure*}[t]
\begin{center}
\includegraphics[width=140mm]{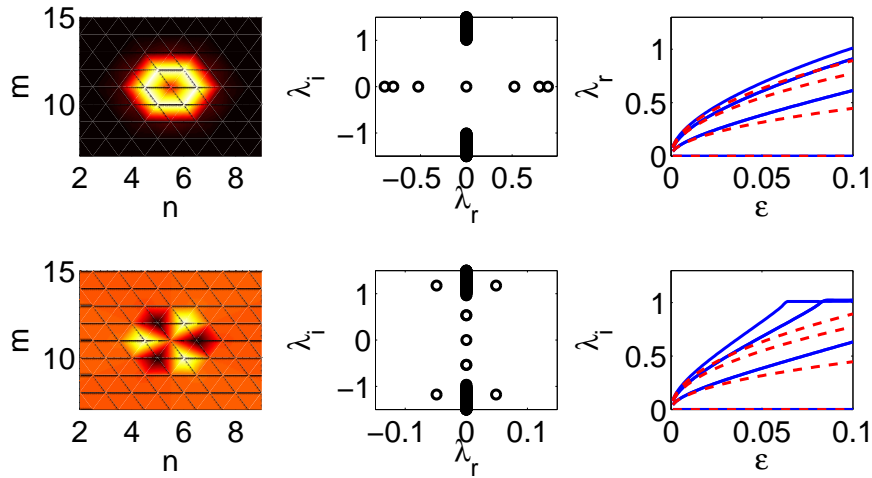}
\end{center}
\caption{(Color online)
Six-site real-valued configurations in a
hexagonal geometry.
The top row corresponds to the {\it unstable}
$\Delta \theta=0$, or ``in phase'' solutions, while the
bottom corresponds to the {\it stable} $\Delta \theta=\pi$, or
``out of phase'' ones.  From left,
the first column is the profile at $\varepsilon=0.08$,
the second column shows
the corresponding linearization spectrum
$(\lambda_r,\lambda_i)$ of the eigenvalues
$\lambda = \lambda_r + i\lambda_i$,
and finally the third column shows the
continuation in $\varepsilon$ of the actual eigenvalues
[real, $\lambda_r$, and imaginary, $\lambda_i$, components] (solid)
and the theoretical predictions given by Eq. (\ref{stab}) (dashed).}
\label{hex_6r}
\end{figure*}

\begin{figure}[!ht]
\begin{center}
\includegraphics[width=80mm]{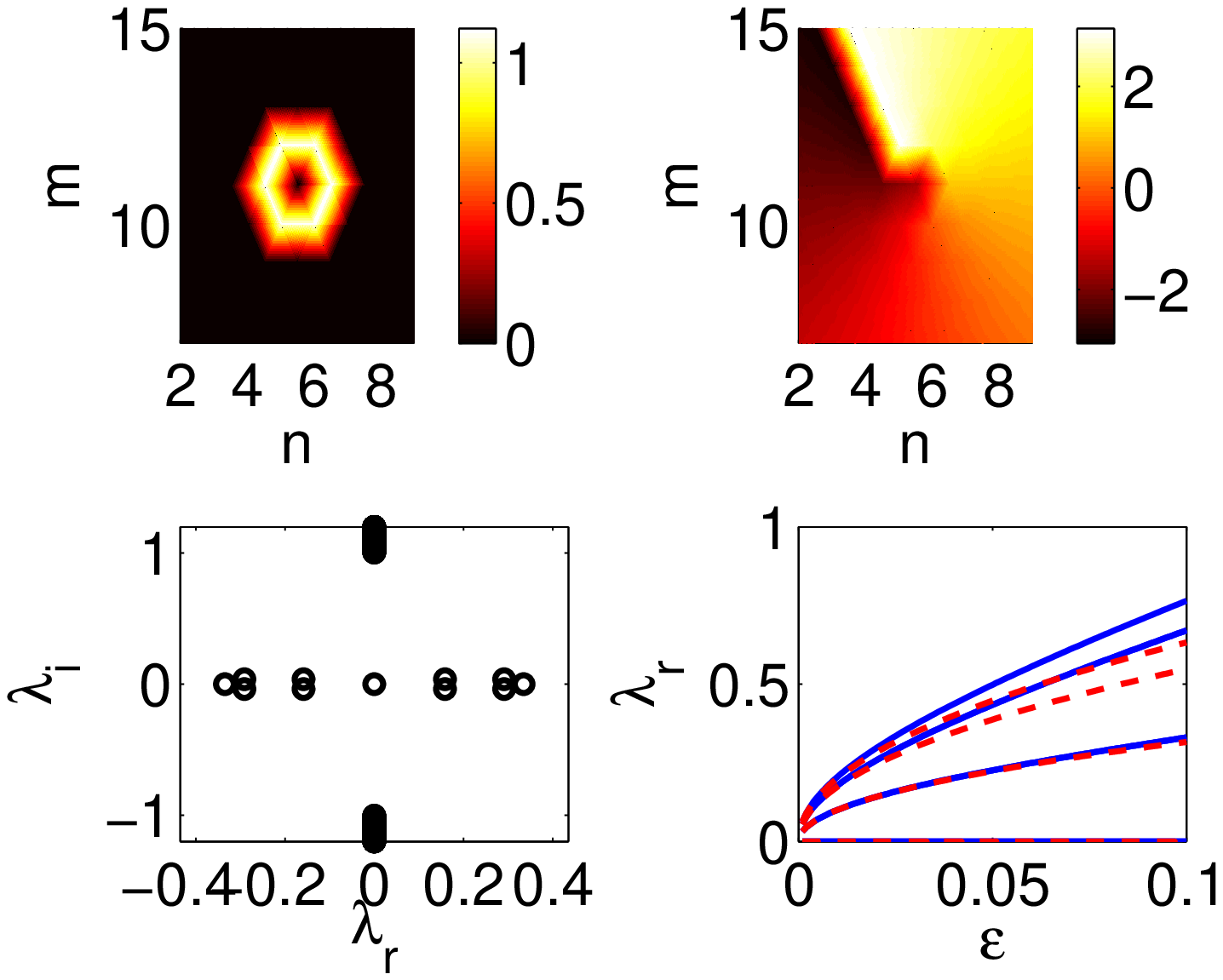}
\includegraphics[width=80mm]{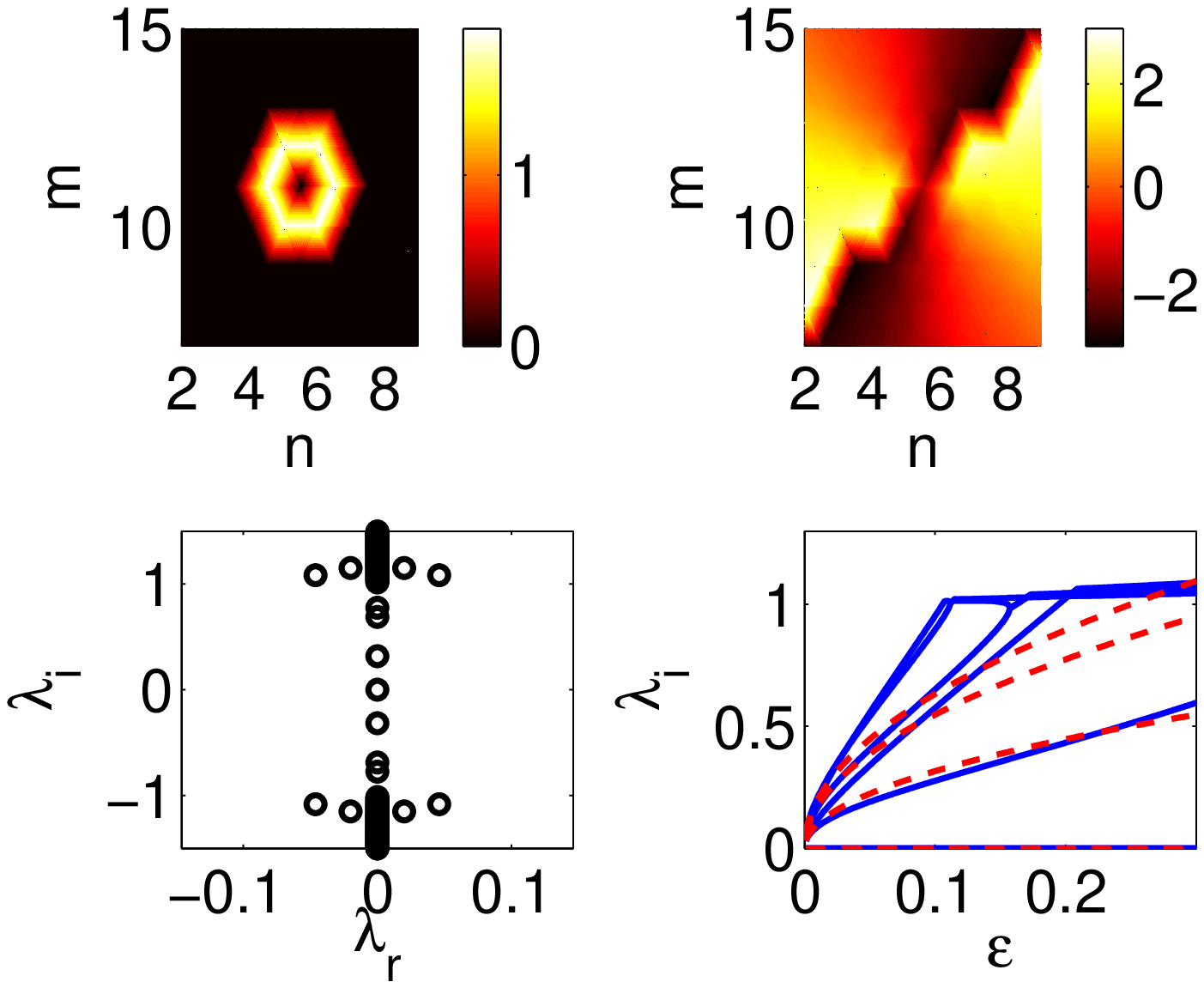}
\end{center}
\caption{(Color online) The six-site vortices in the
hexagonal geometry are shown.
The left four images show the {\it unstable} singly-charged solution branch
with  $\Delta \theta=\pi/3$,
while the right set is for the {\it stable} doubly-charged solution branch with
$\Delta \theta=2 \pi/3$.
The singly-charged vortex is {\it unstable}, while the doubly-charged one
is {\it stable}
(until the oscillatory instability resulting from the
collision of the pairs of negative Krein signature with the
phonon band). The top row
of each set displays the modulus (left) and argument (right) of the solution
with $\varepsilon=0.025$ (top) and $\varepsilon=0.125$ (bottom), while the
bottom left
is the linearized spectral plane and the bottom right is the continuation
of the relevant eigenvalues from the
AC limit, with the solid and dashed lines representing the numerical solution
and the theoretical prediction, respectively.}
\label{hex_6s}
\end{figure}

\begin{figure*}[t]
\begin{center}
\includegraphics[width=140mm]{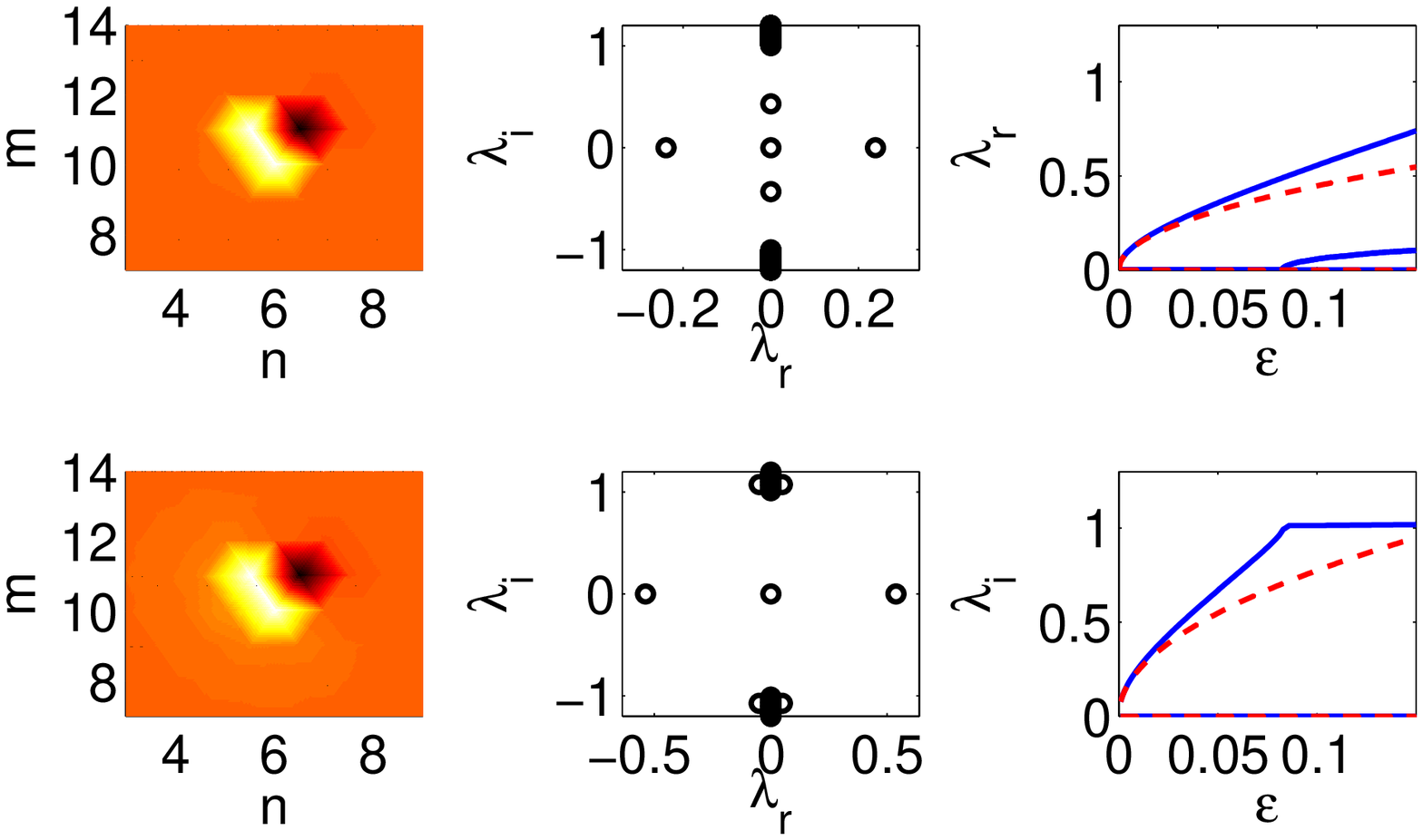}\\
\includegraphics[width=140mm]{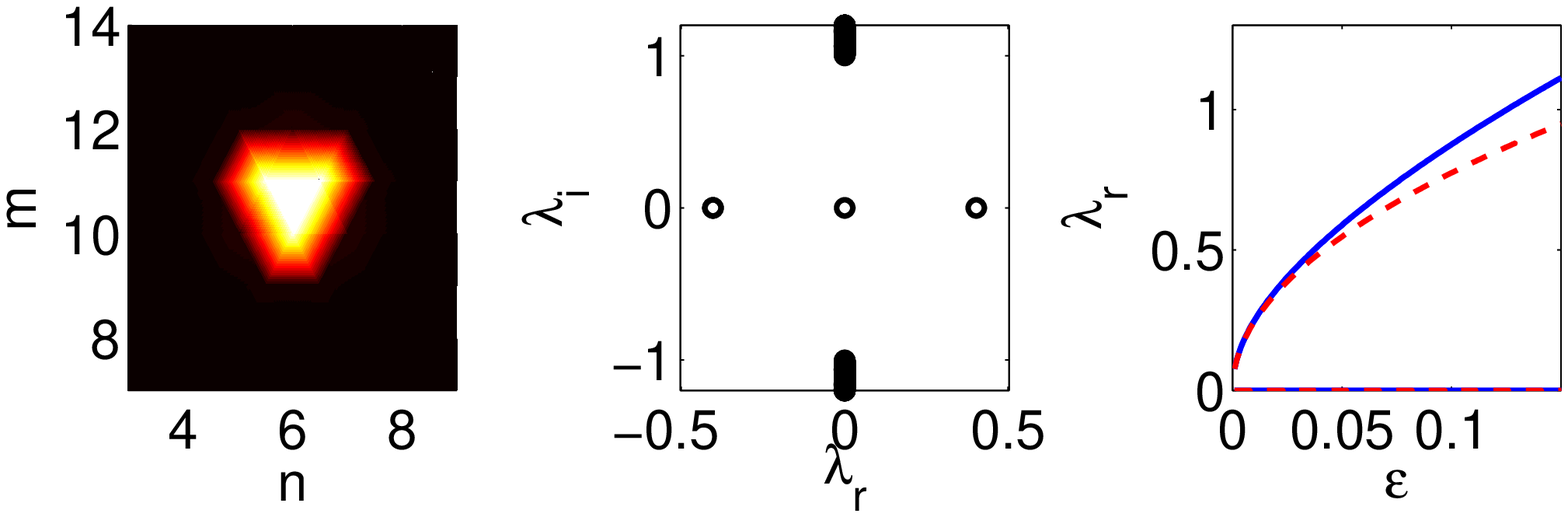}\\
\includegraphics[width=80mm]{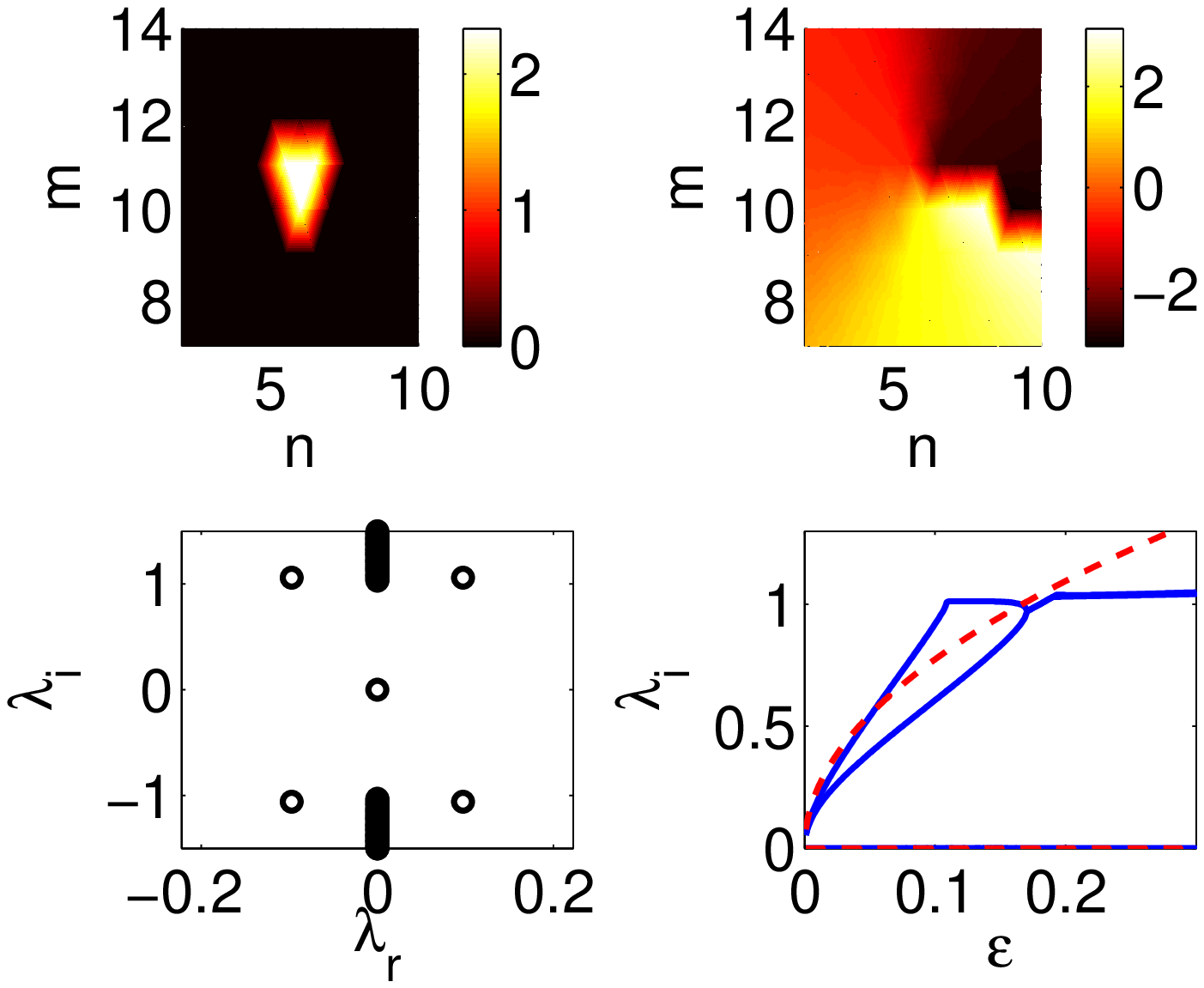}
\end{center}
\caption{(Color online)
The three-site configurations in the hexagonal geometry. The top two rows
present the same panels as Fig. \ref{hex_6r} except for the
{\it unstable} unconventional case where
$\theta_1=\pi$, and $\theta_2=\theta_3=0$ and the third is the
{\it unstable} three-site $\theta=0$ case shown in the same format.
The four panels below that display the {\it stable} (for $\varepsilon \lesssim 0.1$)
three site singly-charged vortex
with $\theta=2\pi/3$. They are (clockwise from top left) the modulus
(for $\varepsilon=0.2$), phase,
continuation of its principal eigenvalues as a function of the
inter-site coupling strength $\varepsilon$,
and linear stability spectrum (for $\varepsilon=0.2$).
Two rows are given for the $0,\pi,0$ case because
one of the null pairs from the AC limit
becomes real (third column of the first row; the
solution and its stability in the first and second column are shown
for $\varepsilon=0.025$), while the other becomes imaginary
(third column of the second row; here the solution and its stability
in the first two columns are for $\varepsilon=0.095$).
}
\label{hex_3}
\end{figure*}


First, we will
study six-site contours, of which we will consider four. The first
two of these are real and are such that either $\Delta \theta=0$ or
$\Delta \theta=\pi$ (any additional combination of $0$ and $\pi$ phases
is also possible but the main qualitative characteristics of stability
will not change from
those reported below).
The relation (\ref{stab}) for $\Delta \theta=0$
predicts double eigenvalue pairs at
$\pm\sqrt{2\varepsilon}$ and $\pm\sqrt{6\varepsilon}$ and single pairs at
$\pm\sqrt{8\varepsilon}$ and $0$, while for $\Delta \theta=\pi$ each of these is
multiplied by the imaginary unity.  Direct numerical computation and
continuation in the coupling parameter $\varepsilon$ from the AC
limit confirm the predictions
presented in Fig. \ref{hex_6r}.  The in-phase configuration with
$\Delta \theta=0$ becomes strongly unstable away from the AC limit, while the
out-of-phase configuration with $\Delta \theta=\pi$ is stable for small
$\varepsilon$.
It should
be noted that more generally any configuration that
has two adjacent in-phase nodes along the six-site contour will also
be unstable for all values of $\varepsilon$,
while the {\it only} potentially stable configuration
of this type (real solution comprising
$0$ and $\pi$ phases) is
the out-of-phase adjacent node structure of  $\Delta \theta=\pi$.
However, even for that configuration, the imaginary eigenvalues which bifurcate
from the origin in the AC limit have the topological property of,
so-called, negative Krein signature \cite{peli2d};
this means practically that they become structurally unstable upon collision
with other eigenvalues, such as
those of the phonon band, which have positive Krein signature.
Hence, when the coupling
becomes sufficiently large ($\varepsilon \gtrsim 0.06$),
these eigenvalues eventually intersect with the
continuous spectrum (the phonon band) edge located
at $\pm i \Lambda$, and result in
Hamiltonian-Hopf bifurcations associated with complex
quartets of eigenvalues and oscillatory instabilities.
Such collisions can be detected in the graphs illustrating the lowest
imaginary eigenvalue parts $\lambda_i$ in the bottom right panel of
Figure \ref{hex_6r}, and are associated with the points where the eigenvalue
trajectories begin to
level out. The spectrum of the solution at $\varepsilon=0.08$
in the bottom middle panel of Fig. \ref{hex_6r} reveals the presence
of such quartets.

Next, we consider the complex valued solutions along the six site
contours, for
which our conditions guarantee
vorticity (i.e., the relevant solutions will be discrete vortices
whose phase completes a round trip of a multiple of $2\pi$ along the
discrete contour).
The fundamental solutions here
are for $\Delta \theta=\pi/3$, which
is a singly-charged vortex,
and $\Delta \theta=2\pi/3$,
which is a doubly-charged vortex. The
relation (\ref{stab}) predicts that the singly-charged vortex will be
unstable with
double eigenvalue pairs $\pm\sqrt{\varepsilon}$ and $\pm\sqrt{3\varepsilon}$,
and single
pairs at $\pm\sqrt{4\varepsilon}$ and $0$. On the other hand, the
doubly-charged vortex
is actually {\it stable} in this lattice geometry for sufficiently small
values of the coupling, with the same pairs as the singly charged vortex,
except multiplied by $i$. Figure \ref{hex_6s} presents both types of
configurations, indeed illustrating
the numerical linear instability
of the former, and numerical linear stability of the latter structures.
Nevertheless, it should be pointed out that for higher values of the
inter-site coupling ($\varepsilon \gtrsim 0.1$)
in this case also, the topological charge $S=2$
solution eventually becomes unstable as well due to oscillatory
instabilities, as is shown in the bottom right continuations of
the relevant eigenvalues of Figure \ref{hex_6s}.
Notice also the generally excellent qualitative and
good quantitative agreement --at least for small
values of $\varepsilon$ (for larger values of the coupling parameter
higher order effects become important)--
between the theoretical predictions of Eq. (\ref{stab}) and
the
numerical
results.

\begin{figure*}[t]
\begin{center}
\includegraphics[width=140mm]{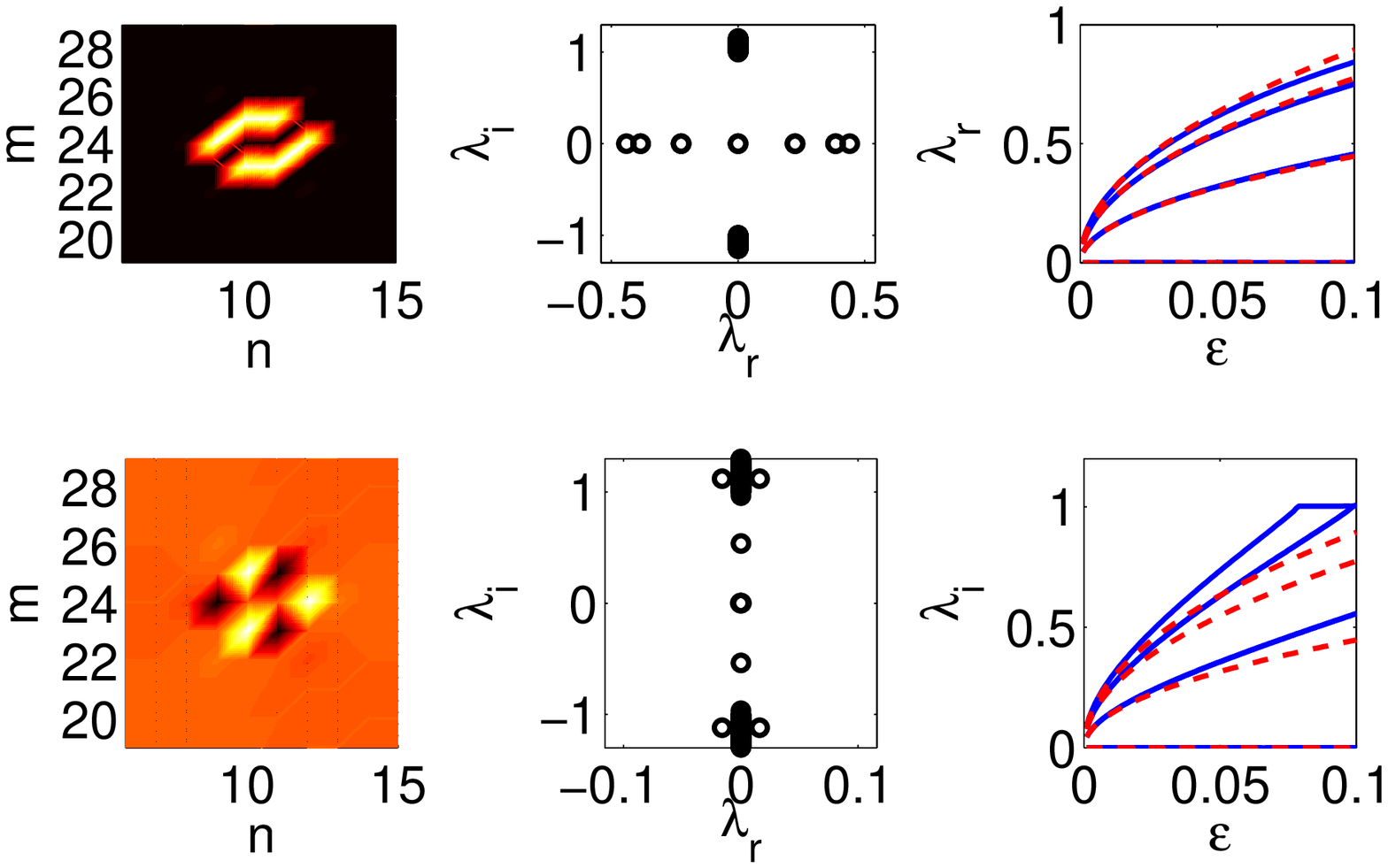}
\end{center}
\caption{(Color online) The same panels as Fig. \ref{hex_6r},
except for the honeycomb geometry.
The particular solutions are for $\varepsilon=0.025$ (top)
and $\varepsilon=0.095$ (bottom).}
\label{honey_6r}
\end{figure*}

\begin{figure}[!ht]
\begin{center}
\includegraphics[width=80mm]{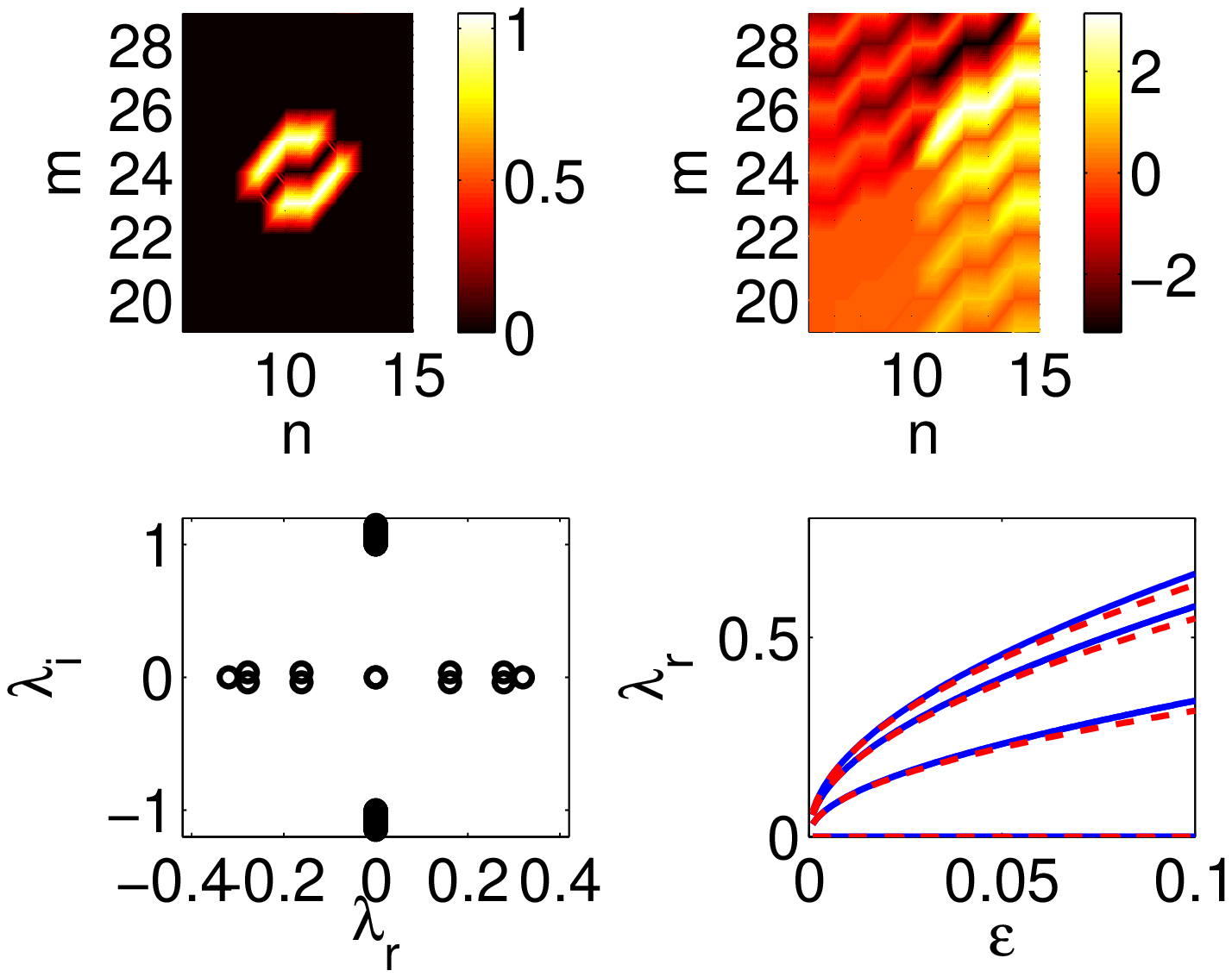}
\includegraphics[width=80mm]{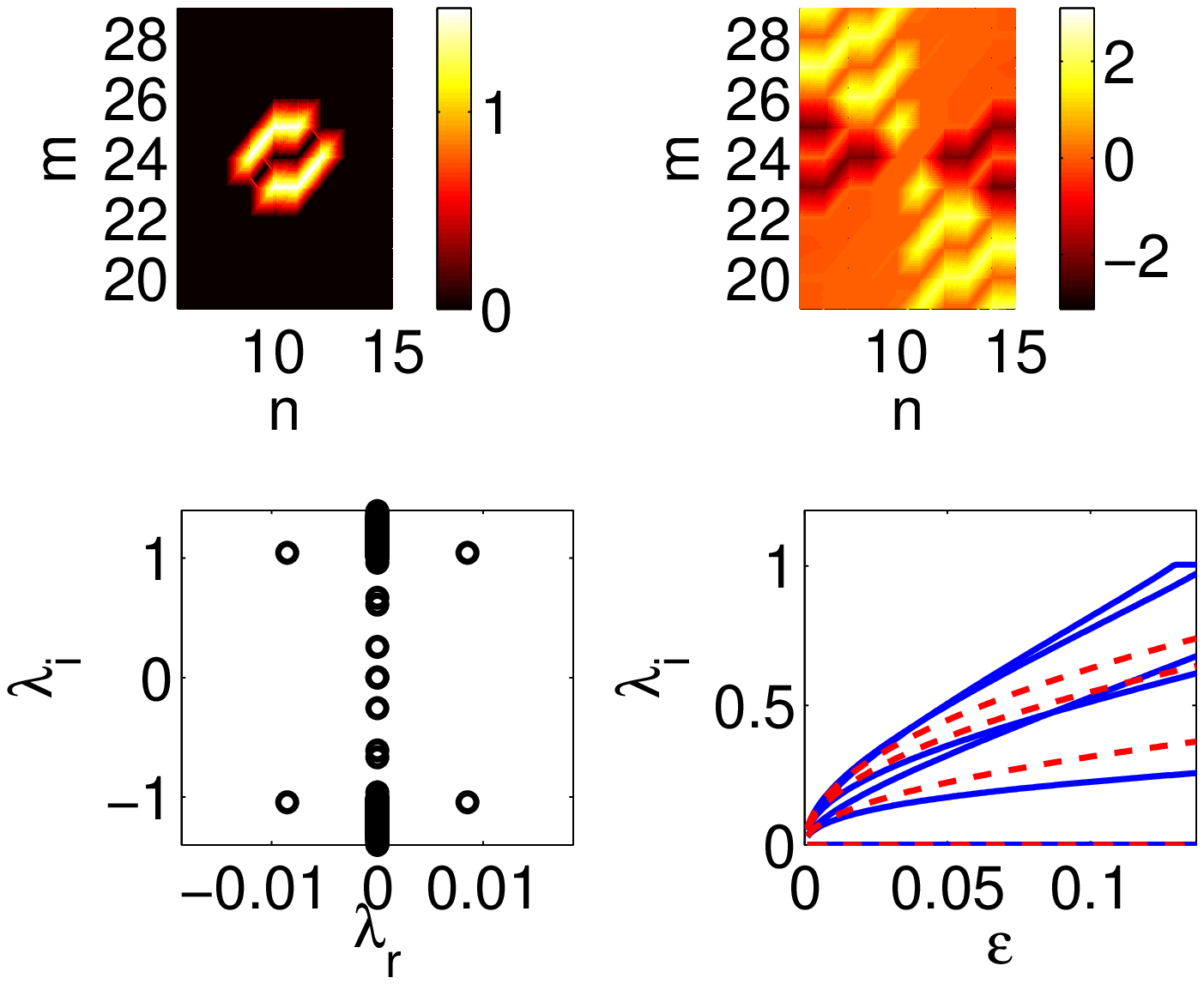}
\end{center}
\caption{(Color online) The same as Fig. \ref{hex_6s}
except for the honeycomb lattice.
The particular solutions given are for the coupling constants
$\varepsilon=0.025$ (left) and $\varepsilon=0.135$ (right).}
\label{honey_6s}
\end{figure}

\begin{figure*}[t]
\begin{center}
\includegraphics[width=140mm]{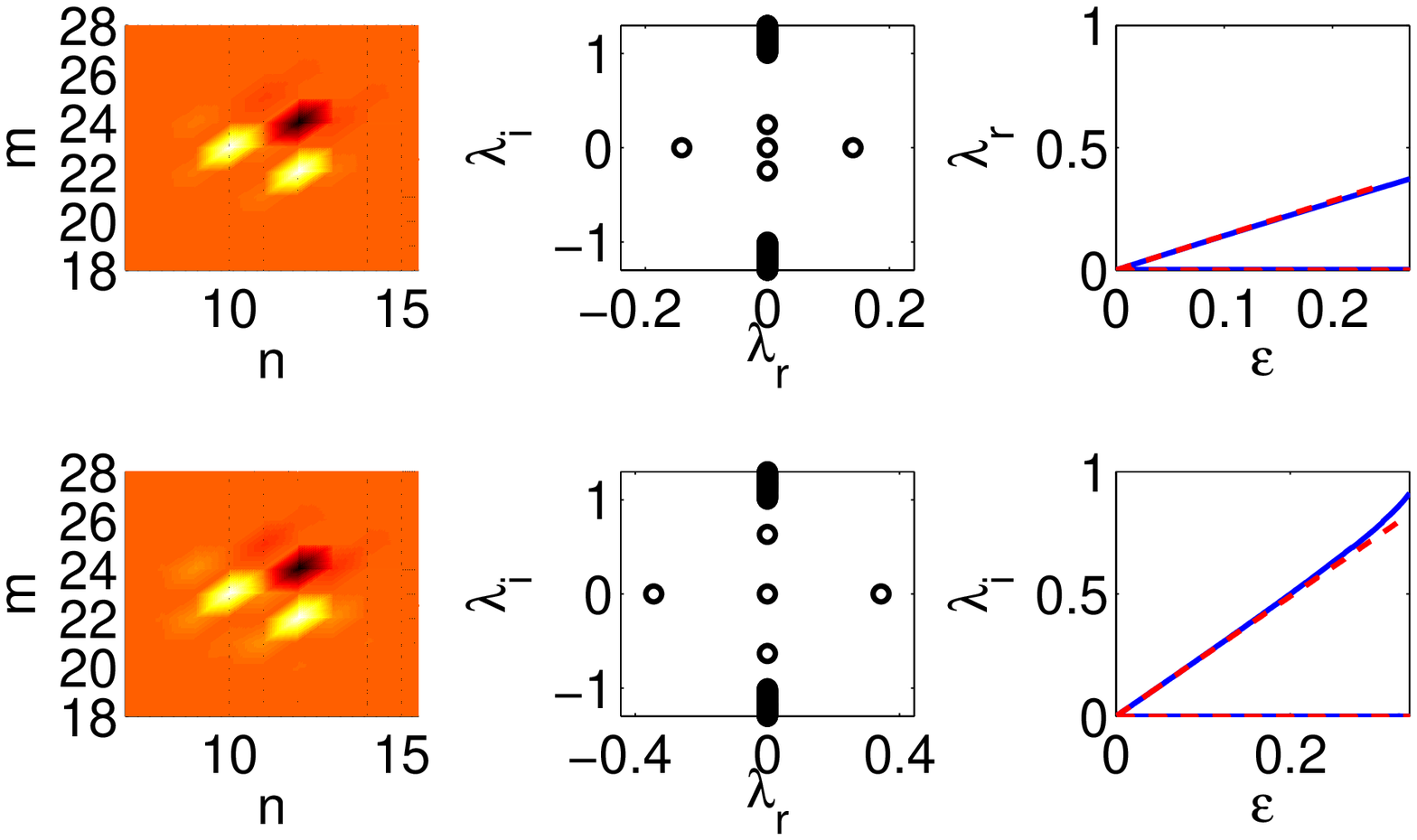}\\
\includegraphics[width=140mm]{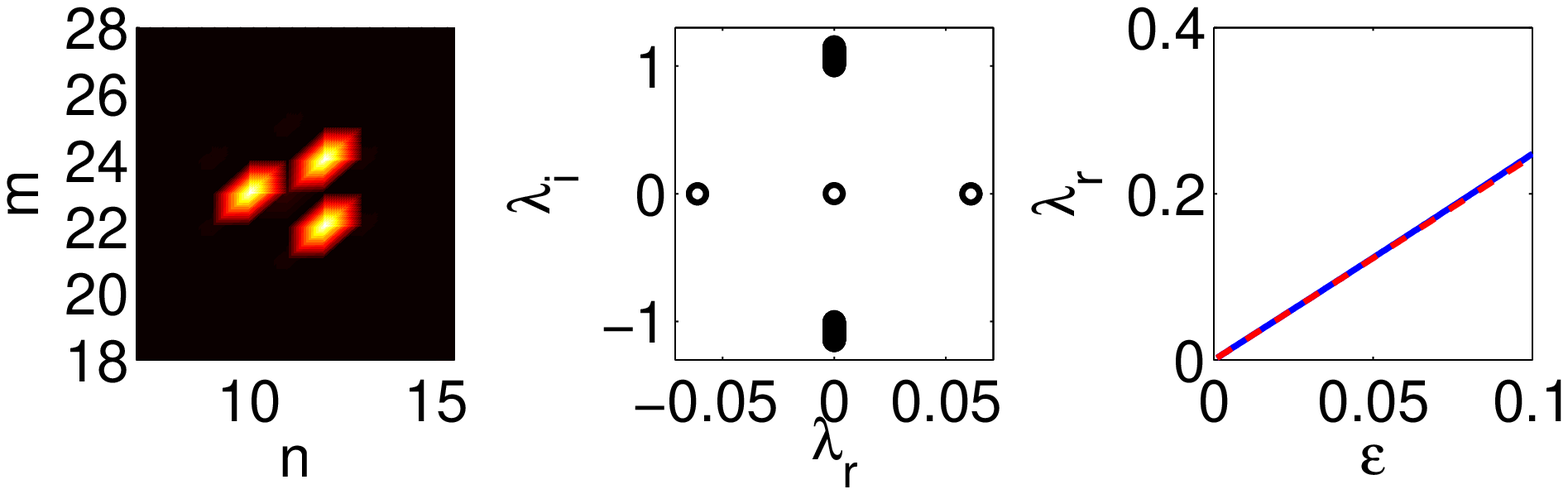}\\
\includegraphics[width=80mm]{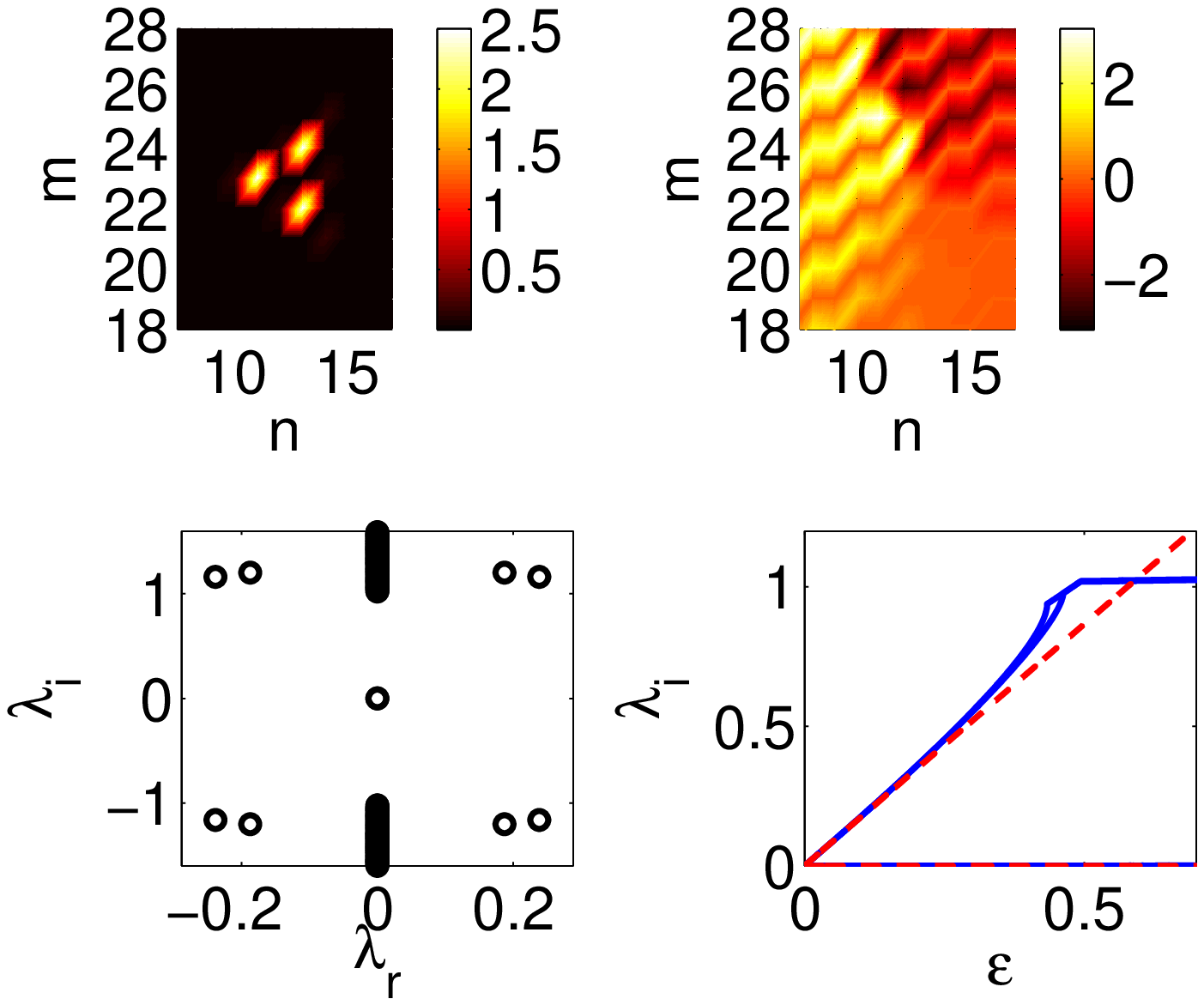}
\end{center}
\caption{(Color online) The same as Fig. \ref{hex_3} except for
the honeycomb lattice geometry. The particular solutions shown
are for $\varepsilon=0.025$ (top and third rows),
$\varepsilon=0.27$ (second row) [the unconditionally {\it unstable}
solutions], and $\varepsilon=0.6$ for charge 1 vortex solution,
which is {\it stable} for  $\varepsilon \lesssim 0.5$, in the
bottom set.}
\label{honey_3}
\end{figure*}

We now turn to the configurations comprised of three lattice
sites. We consider three such cases, similarly to \cite{kouk}
where a Klein-Gordon model was considered.
The first two are the standard real ($\Delta \theta=0$) and
complex-valued ($\Delta \theta=2\pi/3$, corresponding to a discrete
vortex of topological charge $S=1$) ones, while the last one is
the non-standard case with phases $0$, $\pi$ and $0$ for the three
sites. For $\Delta \theta=0$ the theoretically  predicted double pair
of $\pm \sqrt{6\varepsilon}$ and pair at $0$ are confirmed
to exist in the middle row of Figure \ref{hex_3},
while for the discrete singly-charged vortex solution
with $\Delta \theta=2\pi/3$, the double pair
$\pm \sqrt{3\varepsilon}i$ and a pair
at $0$ are also found to reasonably approximate its linearization
eigenvalues in the bottom row of Fig. \ref{hex_3}. It should be noted
that
for larger coupling ($\varepsilon \gtrsim 0.02$) the double pair of
eigenvalues splits,
as can be observed in the numerical results. These solutions also become
unstable as usual after the first eigenvalue collides with the continuous
spectrum at $\varepsilon \approx 0.09$.
On the other hand, for the configuration with phases $0$, $\pi$ and $0$
of the top rows,
one pair of eigenvalues remains at the origin, due to the phase invariance,
while one of the remaining two pairs becomes imaginary
(theoretically predicted as
$\pm i\sqrt{6\varepsilon}$) and the other one becomes real
(predicted as $\pm \sqrt{2\varepsilon}$). We can again observe that
the full numerical results agree well not
only qualitatively but
also even quantitatively with the theoretical description.

\subsection{Honeycomb Geometry}
\label{hon_exstab}

We now explore the same configurations systematically in the case of the
honeycomb lattice geometry, in which each node has three neighbors as opposed
to six. In. Fig. \ref{honey_6r}, we again consider two representative real
configurations, namely the in-phase six-site structure (top row),
and the out-of-phase hexapole, where adjacent neighbors have a relative
phase shift of $\pi$. We find that the principal stability characteristics
are
similar to those in the hexagonal case, in agreement with the theoretical
prediction. In fact, we can observe that quantitatively the agreement
of the linearization eigenvalues is arguably even better between
theoretical predictions and full numerical computations in this case.
This is because the central site, mediating second-order
inter-site interaction between the excited sites, is absent in the six-site
configuration in the honeycomb lattice (contrary to
the case for the hexagonal configuration). This feature reduces the role of
higher order corrections to the theoretical predictions and hence
renders the leading order predictions accurate for wider parametric ranges.
This is a feature that we consistently observe throughout
our honeycomb lattice results.

The six-site discrete vortices are illustrated for this
lattice in
Fig. \ref{honey_6s}. Once again, the interesting
feature of the immediate and generic (i.e., independently of the
precise value of $\varepsilon$) instability of the vortex of topological
charge $S=1$ can be observed, while the vortex of topological charge
$S=2$ is stable for small values of the coupling and is destabilized
for sufficiently large couplings ($\varepsilon \gtrsim 0.13$)
by means of oscillatory instabilities.  Note the larger value of
the coupling necessary for the onset of an oscillatory instability
here as compared to the hexagonal case, presumably a result of the
higher order terms present in the latter due to presence of
the center site.

Finally, the interesting
feature of the three-site configurations in
the honeycomb case is that they now constitute {\it next-nearest-neighbor}
configurations.
As a result, the theoretical prediction that should
be compared to the full numerical results is now $ \propto \varepsilon$
rather than $\propto \sqrt{\varepsilon}$. This is clearly seen to be
consonant with the full numerical findings of Fig. \ref{honey_3},
not only for the strongly unstable (with a double real pair
$\pm \sqrt{6}\varepsilon$) configuration of the in-phase case, or for the linearly
stable (for $\varepsilon \lesssim 0.43$)) vortex case (with a double
imaginary pair $\pm \sqrt{2} i\varepsilon$), but also for the top-row,
$0$-$\pi$-$0$ case of one real ($\pm \sqrt{2} \varepsilon$) pair
and one imaginary ($\pm \sqrt{6} i \varepsilon$) pair of eigenvalues.


\section{Dynamical Evolution Results}
\label{dyno}

\begin{figure*}[t]
\begin{center}
\includegraphics[width=140mm]{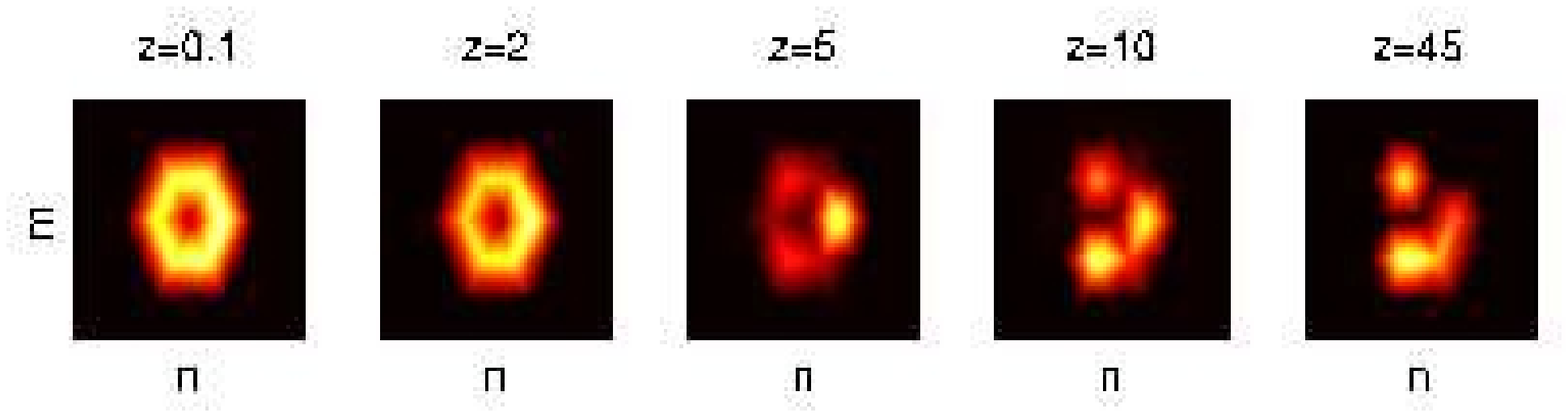}\\
\includegraphics[width=140mm]{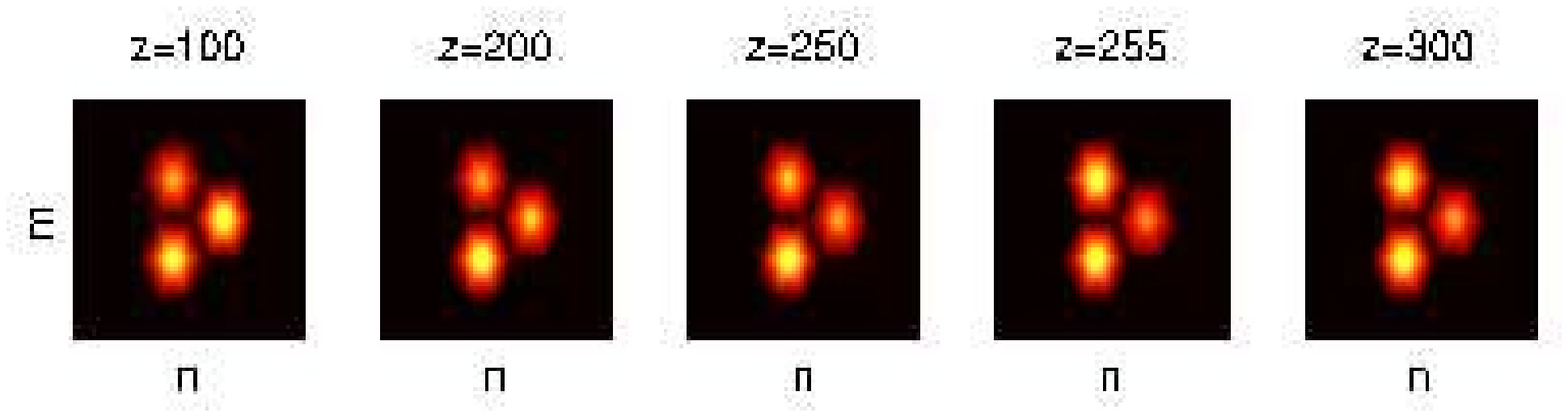}\\
\includegraphics[width=140mm]{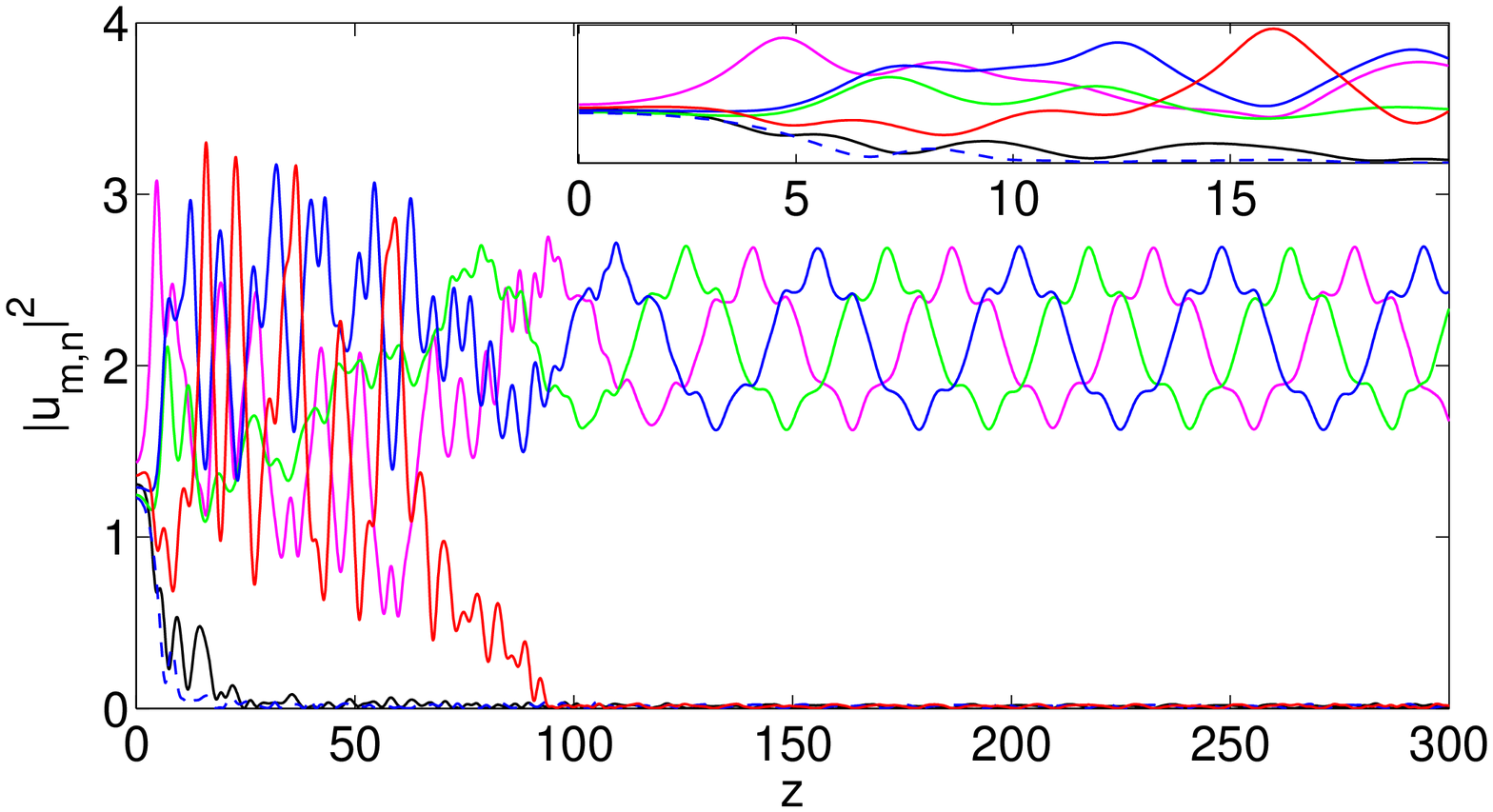}\\
\includegraphics[width=85mm]{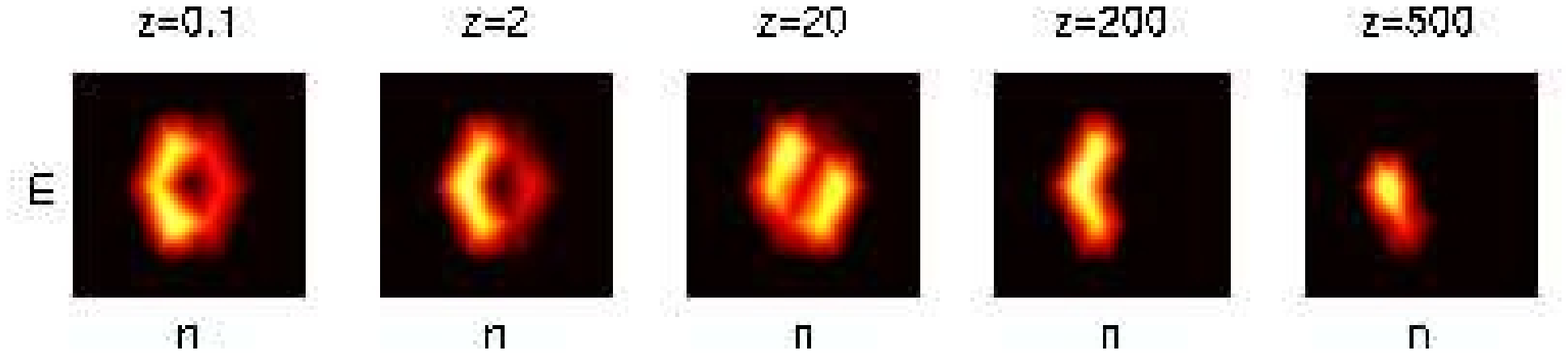}
\includegraphics[width=85mm]{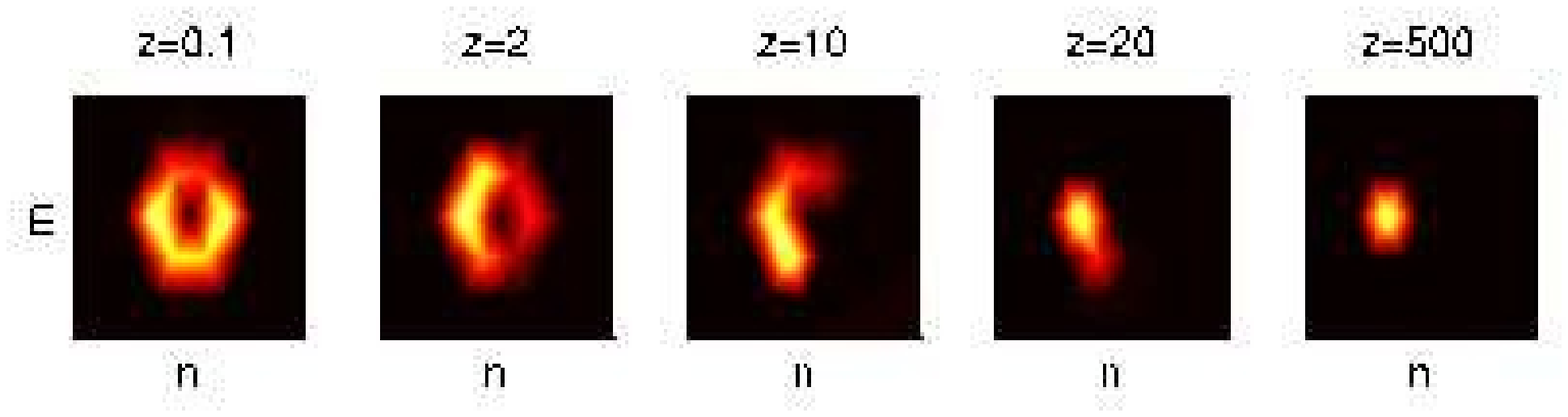}\\
\includegraphics[width=85mm]{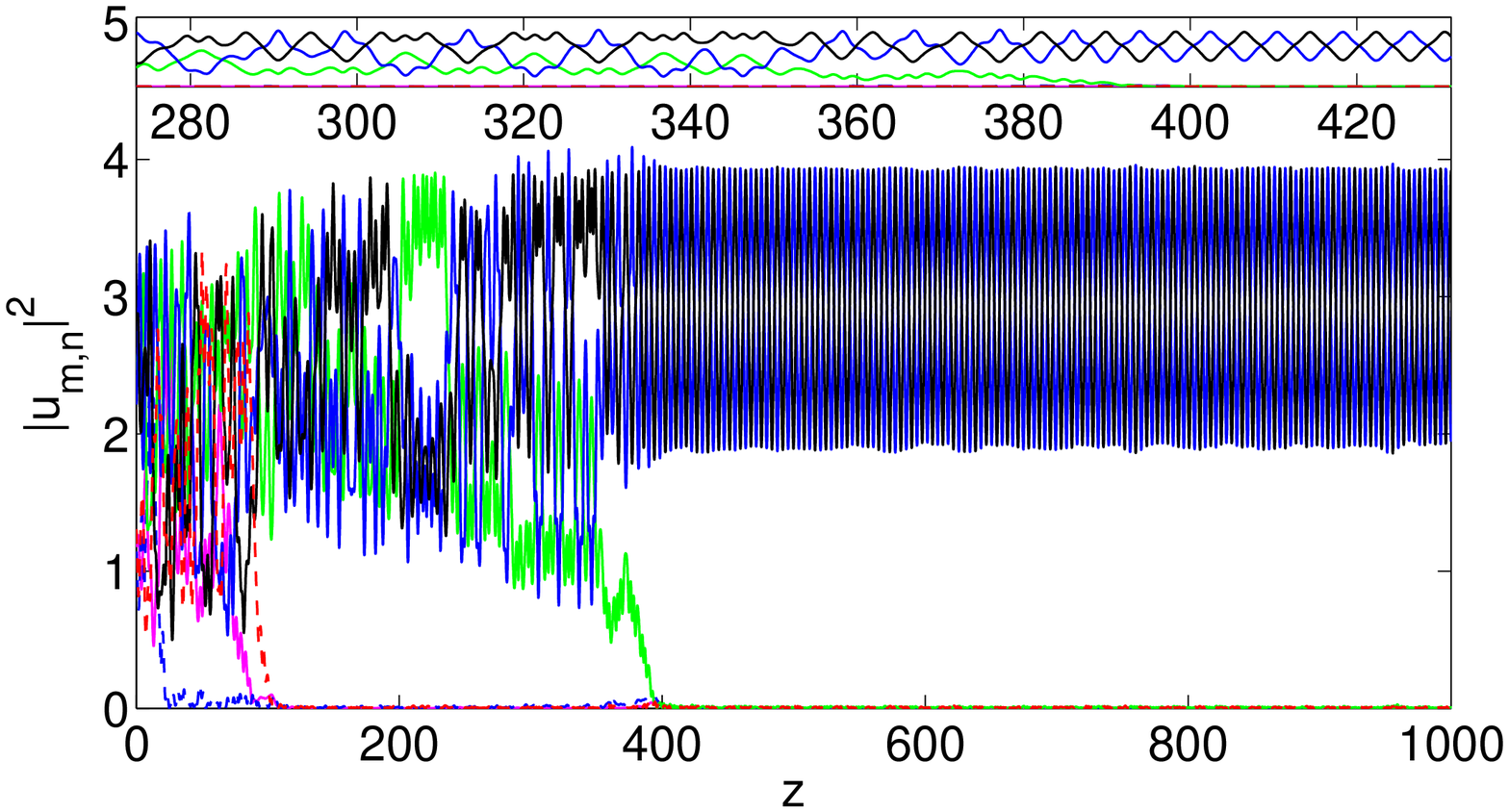}
\includegraphics[width=85mm]{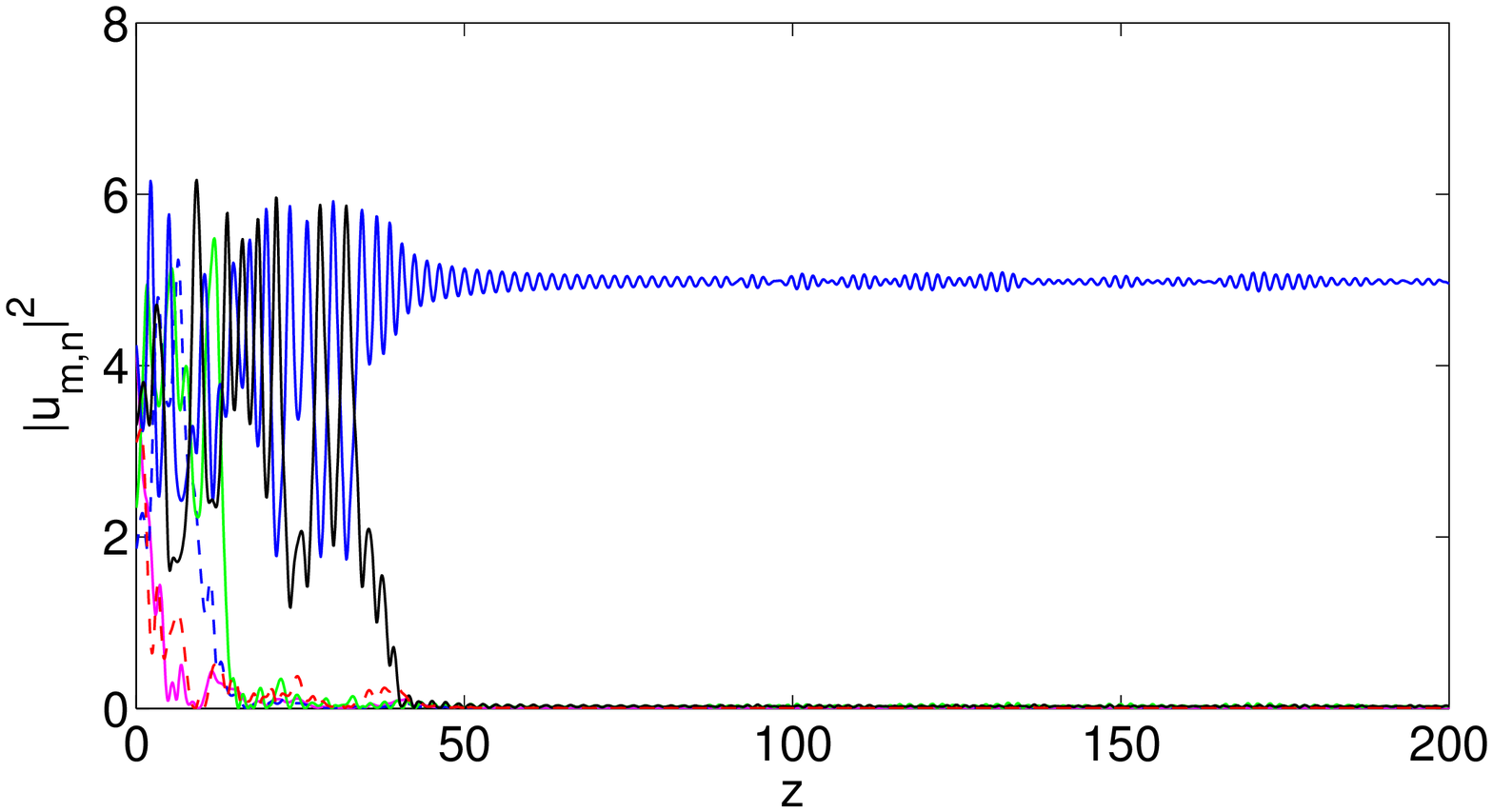}
\end{center}
\caption{(Color online) The top two rows are snapshots in the evolution of the
solution corresponding to $\varepsilon=0.1$ from the family with $\Delta \theta=0$
given in the top row of Figure \ref{hex_6r}, under a small perturbation by a
random noise field to seed the instability. The third row shows the squared
amplitude as a function of propagation distance of the relevant sites,
and the inset shows a closeup of the small distance dynamics. Notice the
structure of the robust three-site periodic structure which emerges after
the original configuration dissolves. Below the third row panel there are
two sets of images for a much larger perturbation of $25\%$ of the intitial
amplitude (left), where the third populated site eventually decays as well
and only two sites persist for long distances, and a much larger
coupling $\varepsilon=0.3$ (right), where the configuration decays
very rapidly and a single site solitary wave remains. There is no clear
correlation between the phases of either of the solutions with
multiple remaining sites.}
\label{hex_6ip_dyno}
\end{figure*}

\begin{figure}[t]
\begin{center}
\includegraphics[width=85mm]{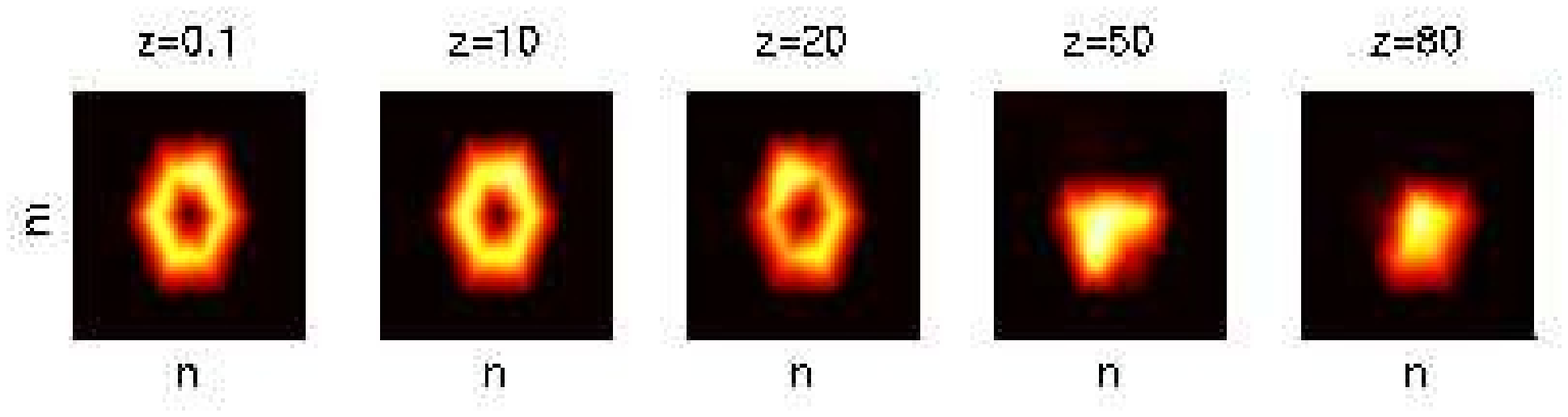}
\includegraphics[width=85mm]{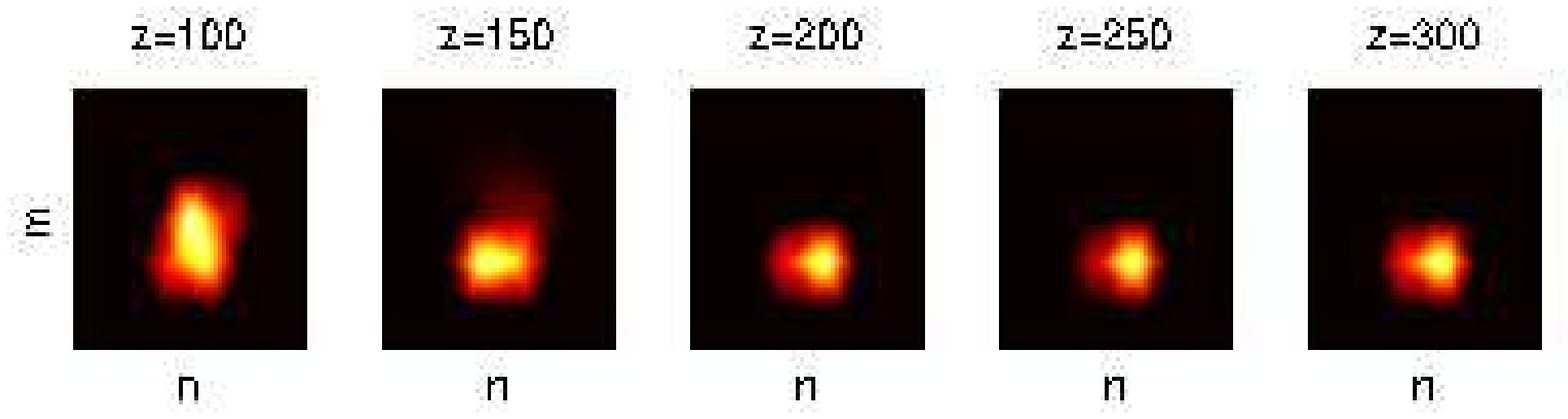}\\
\includegraphics[width=85mm]{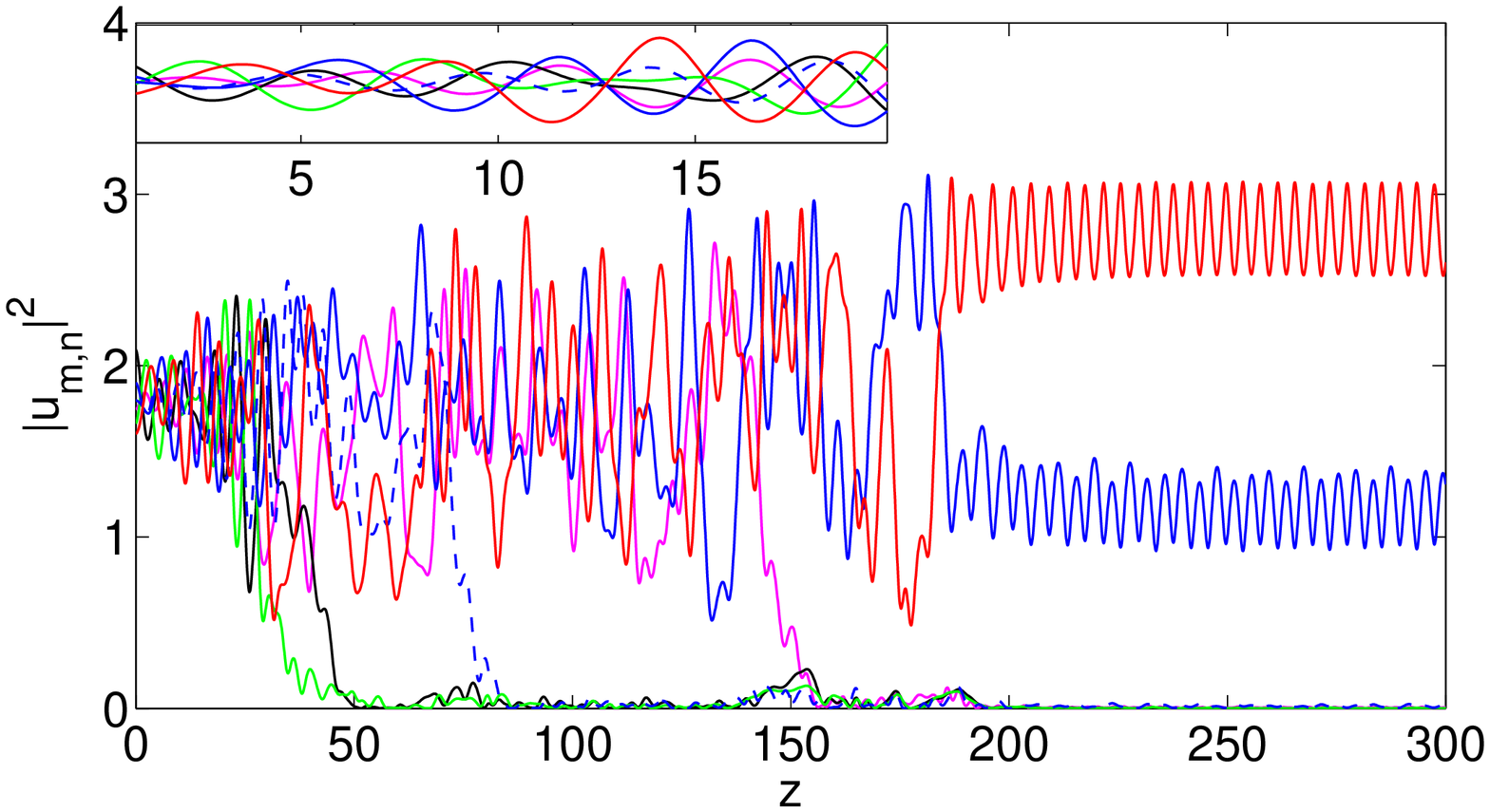}
\includegraphics[width=85mm]{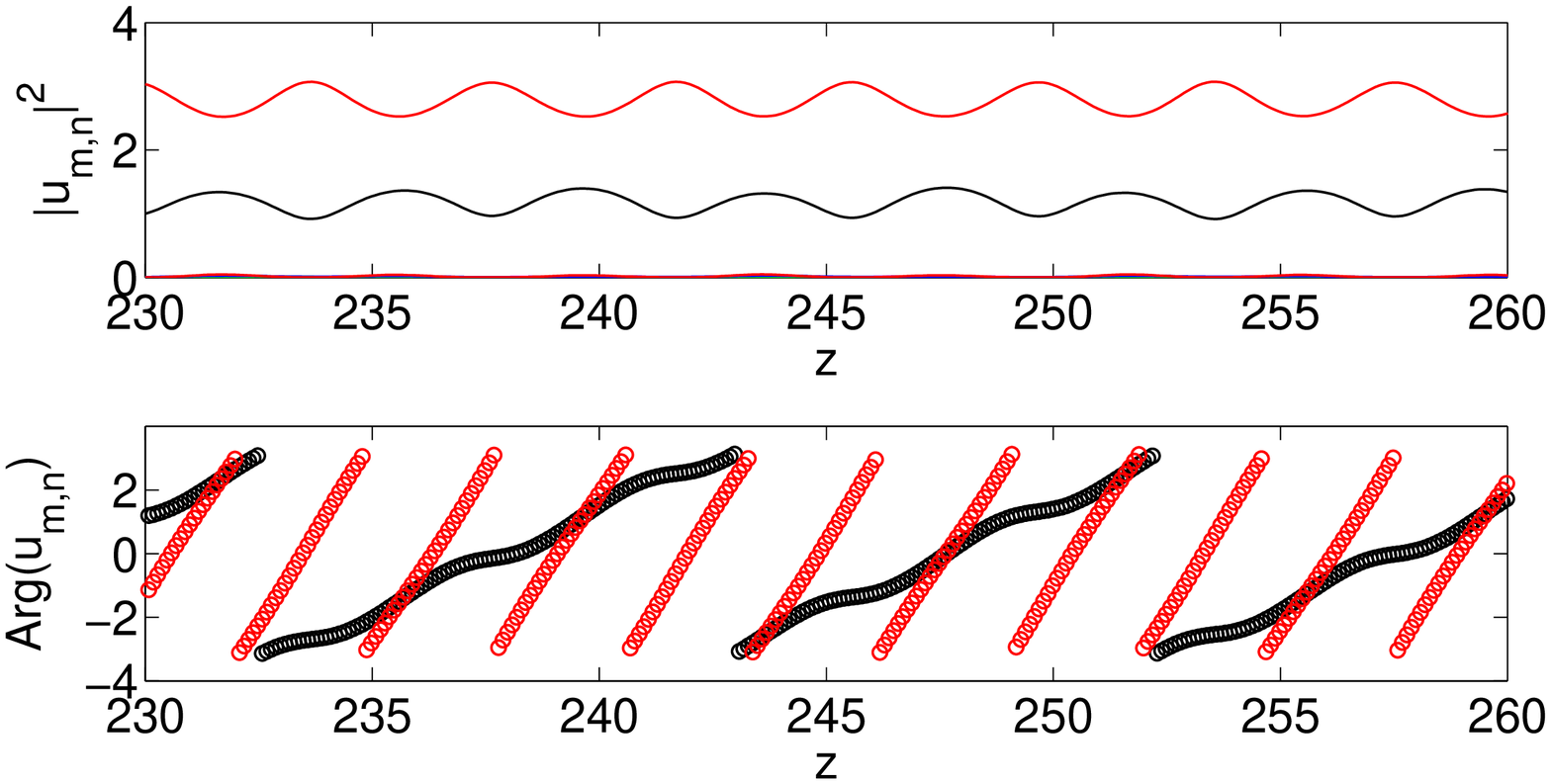}
\end{center}
\caption{(Color online) The same set of figures as in Fig.
\ref{hex_6ip_dyno} (top row, left panels) except
for the solution from the out-of-phase family in the bottom row of
Fig. \ref{hex_6r} are
shown, also for $\varepsilon=0.1$.
On the right
a closeup of the amplitude
oscillations (top) and phase correlation (bottom)
present in the remaining breather is
given.
As one can see, the distance until the initial configuration breaks
down is much longer than for the in-phase case, as expected from the
linear stability analysis (cf. the inset here and in
Fig. \ref{hex_6ip_dyno}).
The ultimate surviving configuration here contains a two-site breathing
structure (see bottom left panel of the squared amplitudes'
evolution), in which there is a
difference in amplitude and the phases of the the two sites
are the same when the amplitudes are closer and
opposite when they are further apart.
}
\label{hex_6op_dyno}
\end{figure}

\begin{figure}[t]
\begin{center}
\includegraphics[width=85mm]{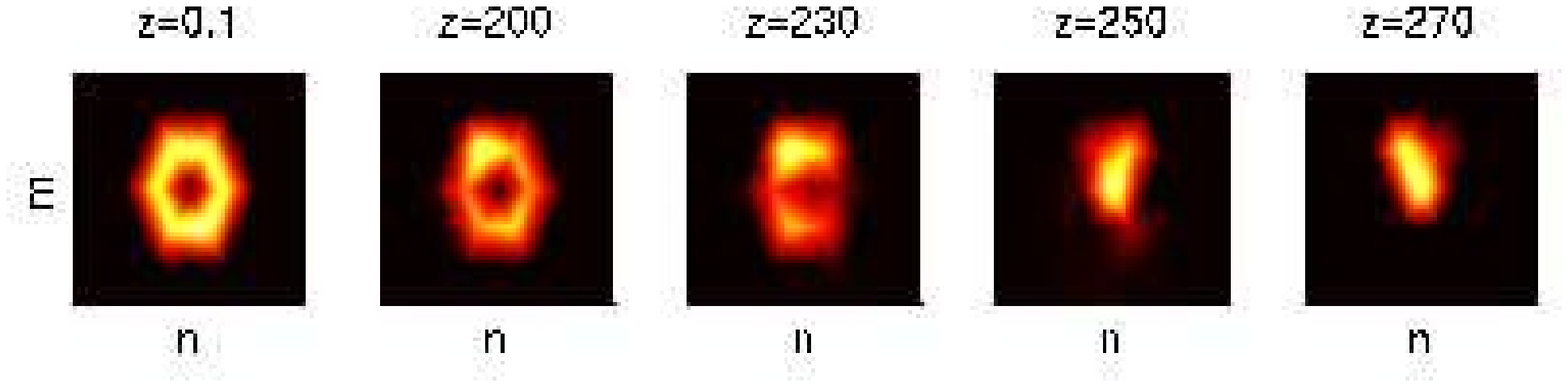}
\includegraphics[width=85mm]{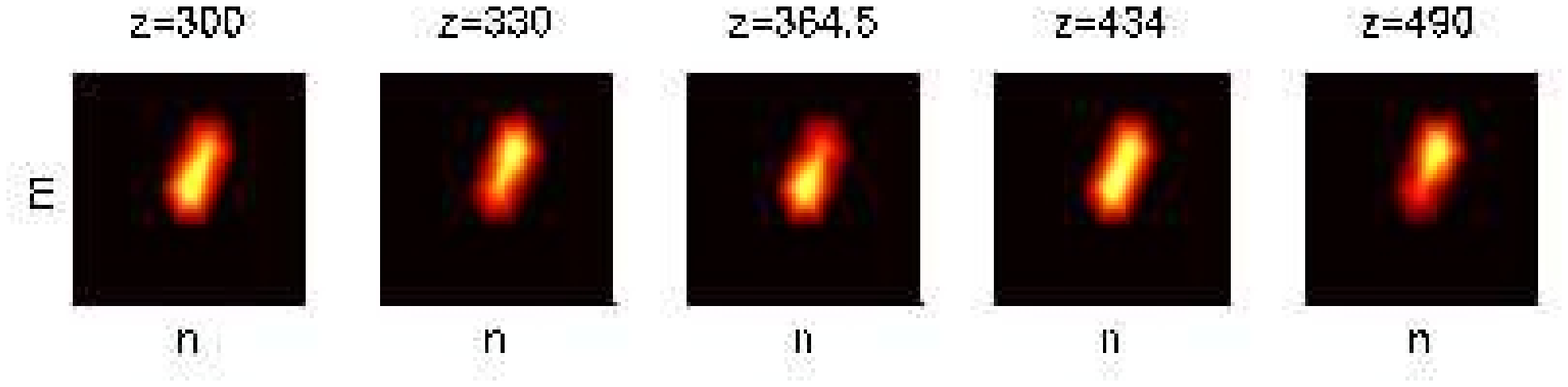}\\
\includegraphics[width=85mm]{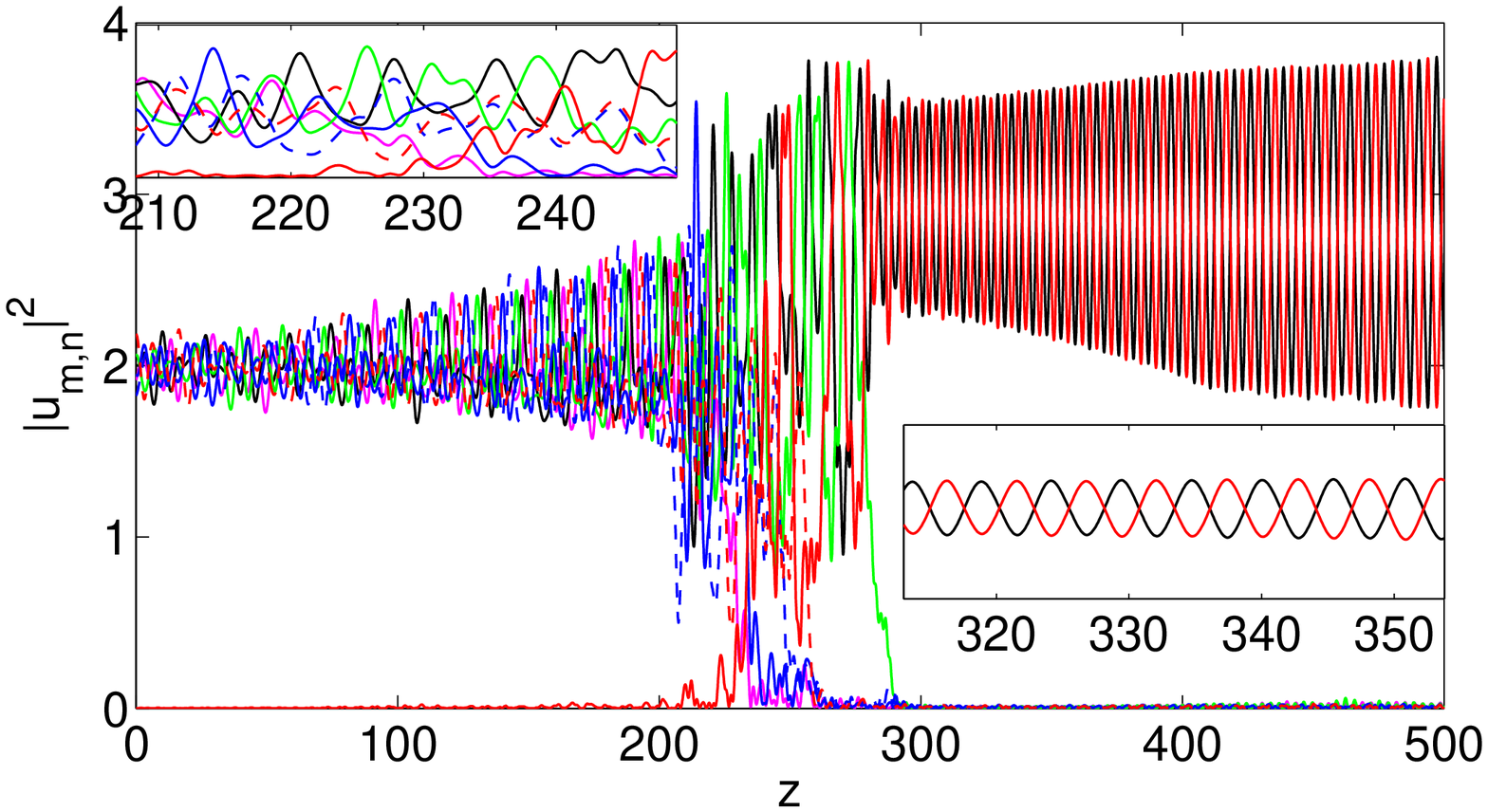}
\includegraphics[width=85mm]{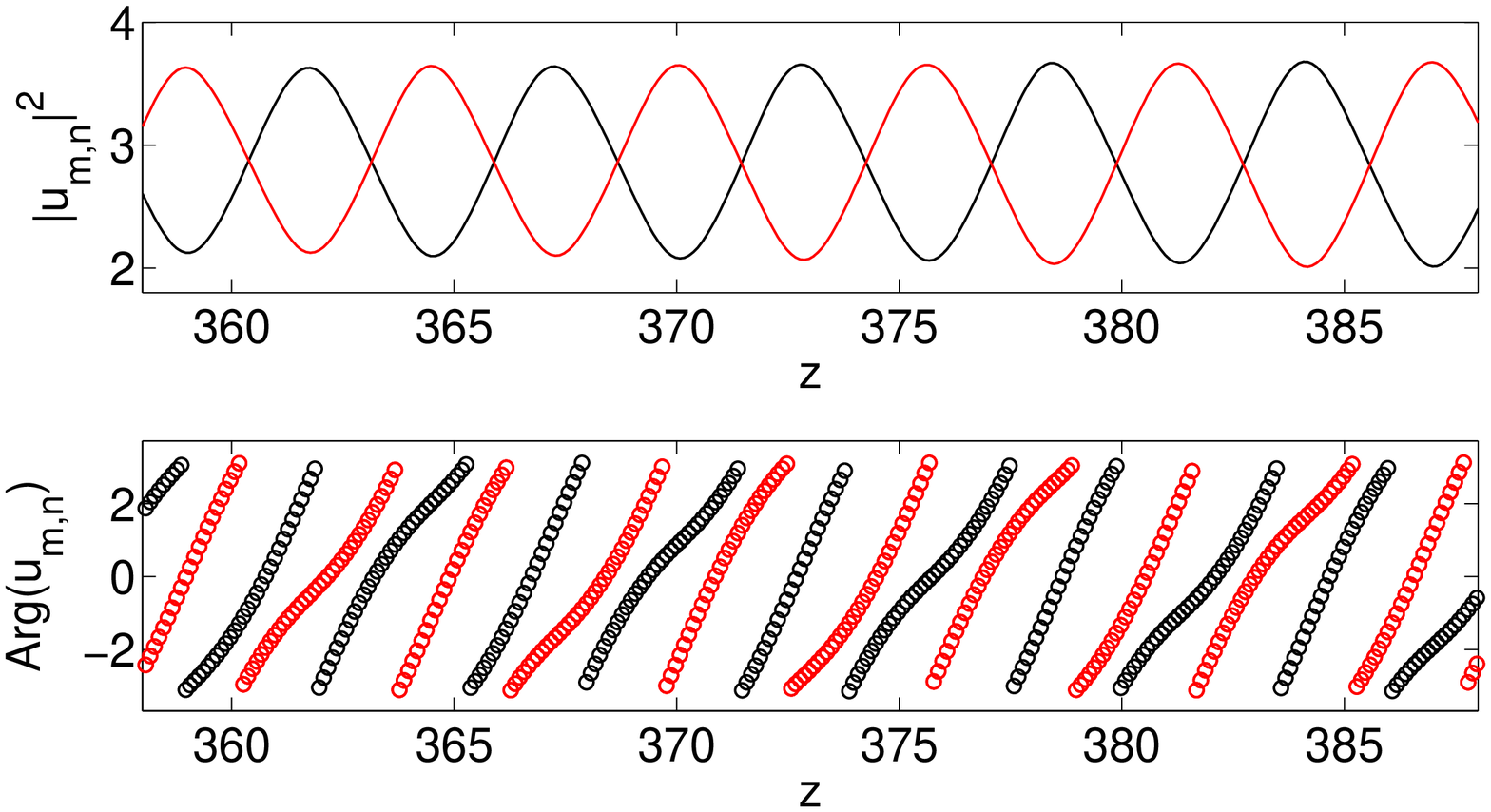}\\
\end{center}
\caption{(Color online)
The same set as the previous images, this time for
a solution from the charge-two family from the
bottom panels of Fig. \ref{hex_6s} with $\varepsilon=0.125$,
i.e., large enough that a quartet of eigenvalues emerges. The long
distance until initial breakup confirms the linear stability analysis,
but a
two-site
(including the initially unpopulated center site)
breathing structure persists after the disintegration of the initial
structure. The inset panels show closeups of the amplitudes
for shorter and longer
propagation distances, respectively. In the right panels, one
can observe that these two sites remain usually out of phase as their
amplitudes oscillate.}
\label{hex_6s2_dyno}
\end{figure}


\begin{figure}[!ht]
\begin{center}
\includegraphics[width=85mm]{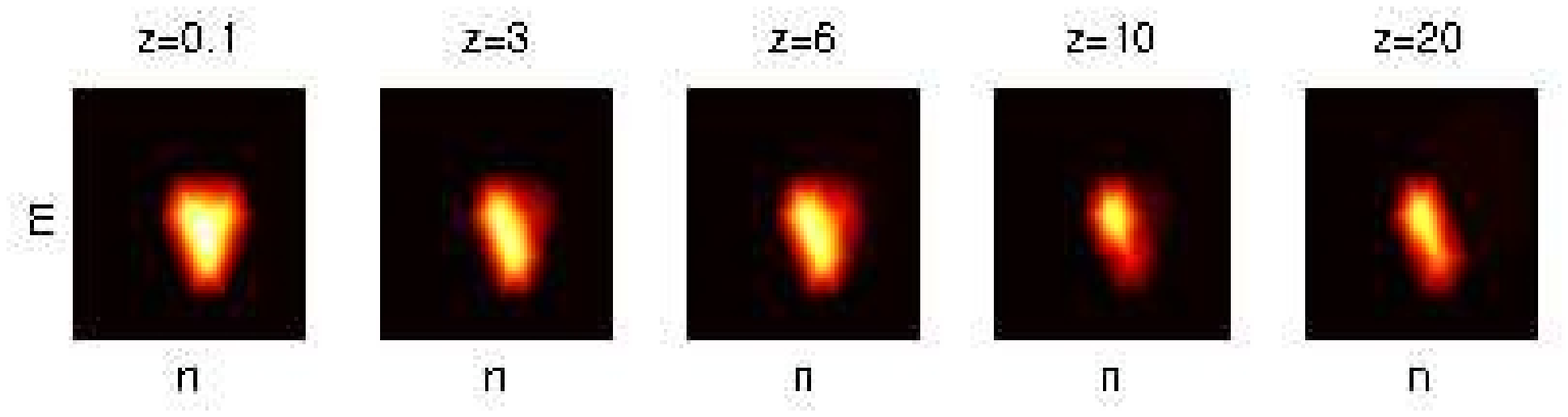}
\includegraphics[width=85mm]{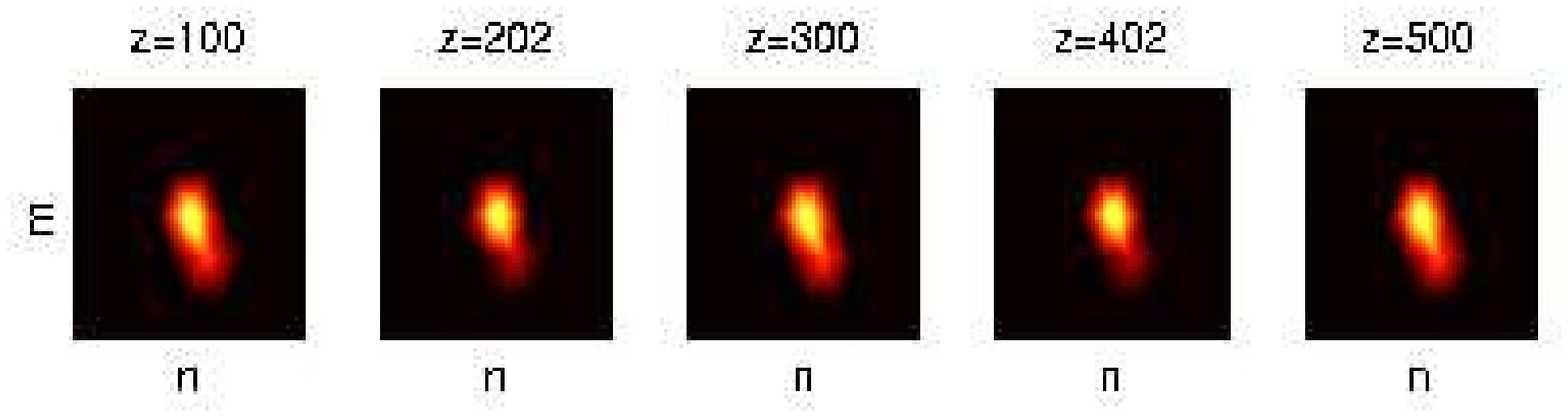}\\
\includegraphics[width=85mm]{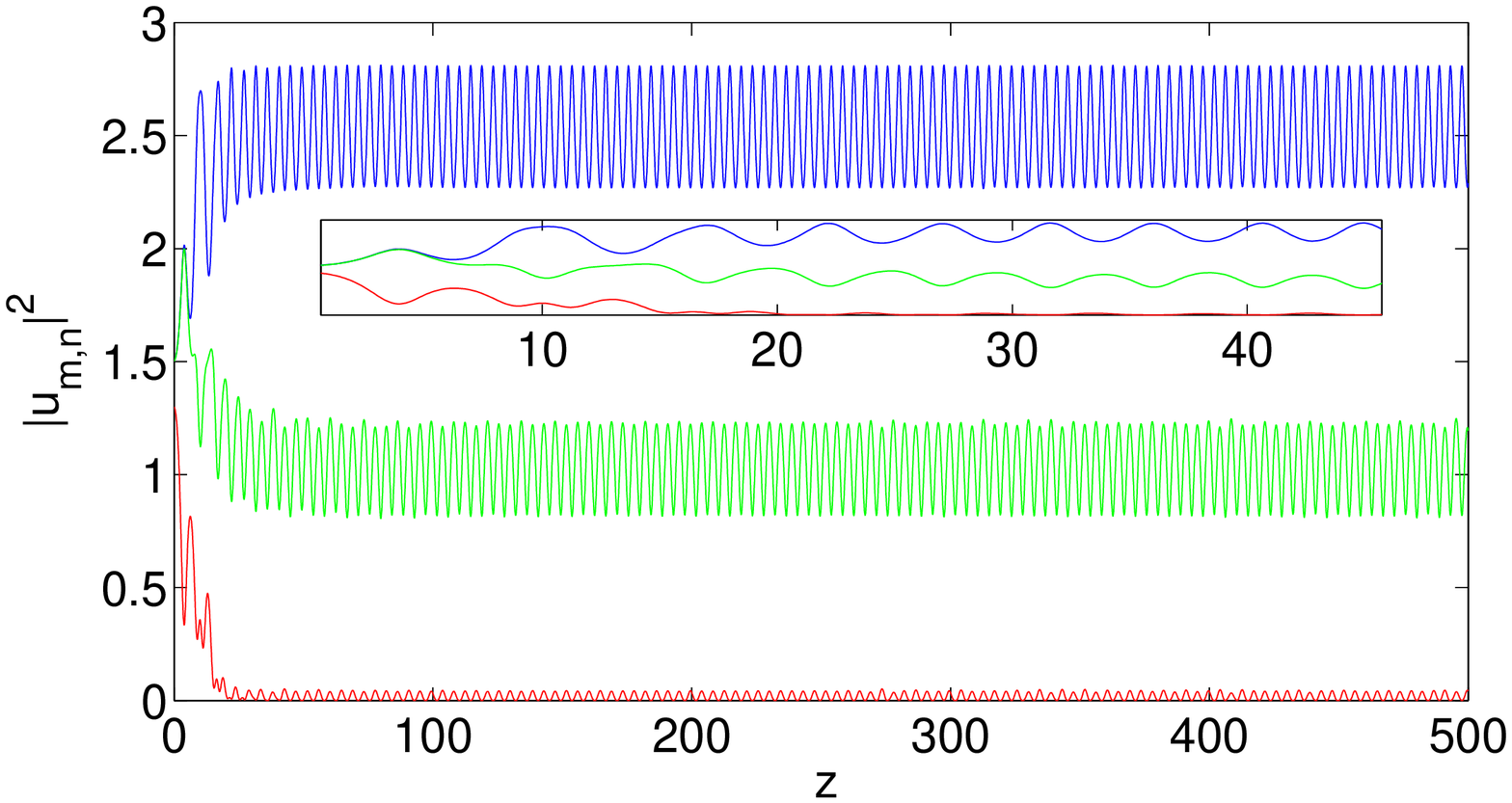}
\includegraphics[width=85mm]{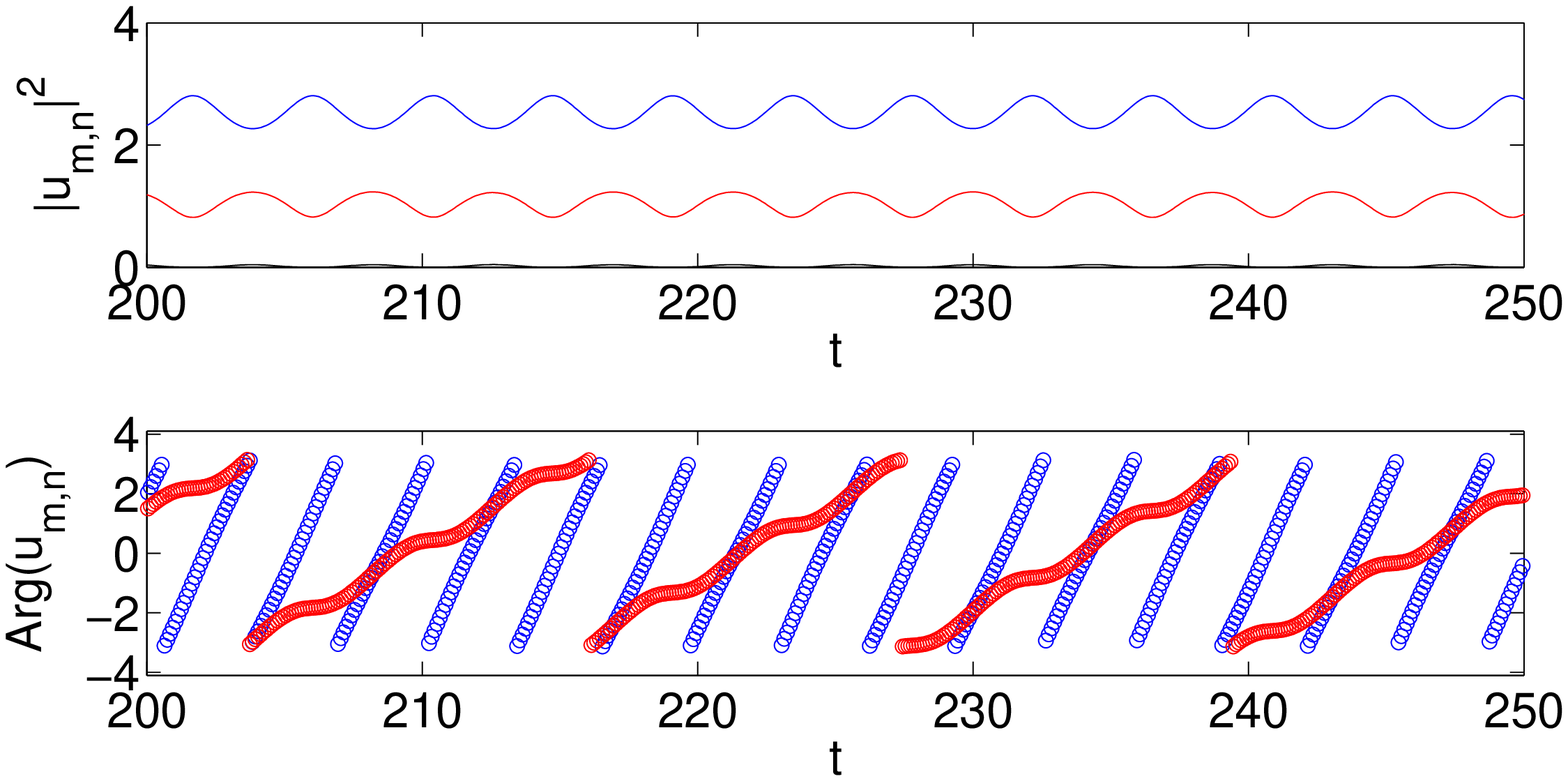}
\end{center}
\caption{(Color online) The in-phase, three-site configuration
from the third row of Fig.
\ref{hex_3} is shown, in which a two-site breather persists
for long propagation distances. The phases are correlated like
the other unequal amplitude two-site breather which resulted
in the evolution of the out-of-phase hexapole shown in Fig.
\ref{hex_6op_dyno}, in which they become in-phase and out-of-phase
depending on whether their amplitudes are similar or considerably different,
respectively.}
\label{hex_3uuu_dyno}
\end{figure}

We now examine the nonlinear dynamics of an unstable solution of each
configuration upon integration of a slightly perturbed waveform
$u=u^s(1+u^r)$, where $u^s$ is the complex valued vector field which
is a stationary unstable solution to Eq. (\ref{dnls_eq1}), and $u^r$ is a
random noise field (i.e., a field in which every entry is a random
variable distributed uniformly in the interval between
$\pm 0.05 {\rm max}_{\{m,n\}}[|u_{m,n}(t=0)|^2]$). Since the
coupling sensitively affects the dynamics
(in particular, larger
coupling facilitates communication between sites and hence propagation
of the instability), we use a fixed
coupling of $\varepsilon=0.1$ for all solutions except for those which are
stable until larger values of $\varepsilon$. In a few seemingly
counterintuitive cases we examine the cases of larger perturbation
and coupling, and find that these effects (more so the coupling) do
indeed influence the dynamical evolution. We will see that several
cases degenerate to similar two-site breathing structures with phase
correlation which may be either out-of-phase
or oscillating
between in- and out-of-phase.

\subsection{Hexagonal Geometry}



First, we explore the
evolution of characteristic unstable solutions
from the families of configurations in a hexagonal geometry given in section
\ref{hex_exstab}.
Within this class we begin with the six-site
configurations. The evolution of the real valued
solution with $\Delta \theta=0$
from the family in the top row of Fig. \ref{hex_6r}
is displayed in Fig. \ref{hex_6ip_dyno}.
The rapid destruction of the original configuration confirms the linear
stability analysis, which predicts strong instability from five pairs of real
eigenvalues. However, for $\varepsilon=0.1$ (top set) and a $5\%$
perturbation, after the destruction of the initial configuration, a
robust
three site oscillating breather state emerges
(note the plot of individual site amplitudes as a function of propagation
distance in the third row of Fig. \ref{hex_6ip_dyno}).
Despite the apparent coherence of the amplitude oscillations, the
relative phases of the sites appear to be uncorrelated and are not shown.
A similar phenomenon is observed for a much larger initial perturbation of $25\%$ of
the initial amplitude (bottom left panels of Fig. \ref{hex_6ip_dyno}),
although here the amplitude oscillations remain
irregular even with three populated
sites, and after a long
distance, a nonlinear dynamical structure emerges in
the form of a two-site breather. Again, however, there is no definite
pattern in their relative phases.
For a much larger coupling, on the other
hand, as shown in the bottom right panels, all sites except for one
decay very rapidly and a single site survives for long distances.
Dynamical evolution of a real valued solution from the bottom row of Fig.
\ref{hex_6r} with $\Delta\theta=\pi$ is displayed in Fig. \ref{hex_6op_dyno}.
The original configuration takes considerably longer to decompose than the
in-phase counterpart given above, confirming the expectation based on the
small magnitude complex quartet of unstable eigenvalues
of the linearized system.
Once again, in this case
a two-site structure with oscillating amplitudes
persists long after the original break-up.  However,
in this case, there is a strong phase correlation,
and when the amplitudes of these sites are close
they are in-phase, while when they
are distant
they are out-of-phase (shown in the right panels).

Next,
we consider the vortex solutions with six sites.
Both of these configurations confirm again the linear stability analysis,
and also both feature two-site breathers for long distances.
The
singly-charged vortex ($\Delta \theta=\pi/3$)
from the left panels of Fig. \ref{hex_6s} decays into a breather
with uncorrelated phases, similarly to the bottom left panel of Fig.
\ref{hex_6ip_dyno} and, hence is not shown.
The evolution of the more stable doubly-charged vortex
($\Delta \theta=2\pi/3$) from the right panels of
Fig. \ref{hex_6s} is in Fig. \ref{hex_6s2_dyno}.
Notice the almost harmonic oscillations of the breather shown in the inset
for the doubly-charged case.
Here the two sites are also of comparable amplitudes, but
as they oscillate they remain usually out-of-phase with each other as
shown in the right panels.
Another feature of both of
these cases is that one of the two ultimately
surviving sites is the originally
unpopulated center site, which inherits mass from
other sites when they decay
(see also the insets in each figure).

We now consider the three-site configurations.
Both the $\theta_i=0,\pi,0$ solution from the
top rows of Fig. \ref{hex_3} (unstable
due to one real pair of eigenvalues)
and the more unstable $\Delta \theta=0$ solution given below that
(which is unstable due to two pairs of real eigenvalues), ultimately
decay into in-phase/out-of-phase breathers such as the one shown in
Fig. \ref{hex_6op_dyno}, although faster in the latter case due to the two
unstable directions.  The latter is displayed in Fig. \ref{hex_3uuu_dyno}.
The final three site configuration is from
the potentially stable $\Delta \theta=2\pi/3$
family in the bottom panels of Fig. \ref{hex_3}.
The imaginary
eigenvalues with negative Krein signature do not reach the continuous
spectrum until a large coupling value in this case, and so
we investigated the dynamics for $\varepsilon=0.2$.
Despite the magnitude of the
growth rate being comparable with the previous cases, two
of the original populated sites here rapidly decay and a
robust single site remains.
This may be a result of the stronger site
interaction induced by the larger coupling.

\subsection{Honeycomb Geometry}
\label{dyno_honey}

\begin{figure*}[!ht]
\begin{center}
\includegraphics[width=85mm]{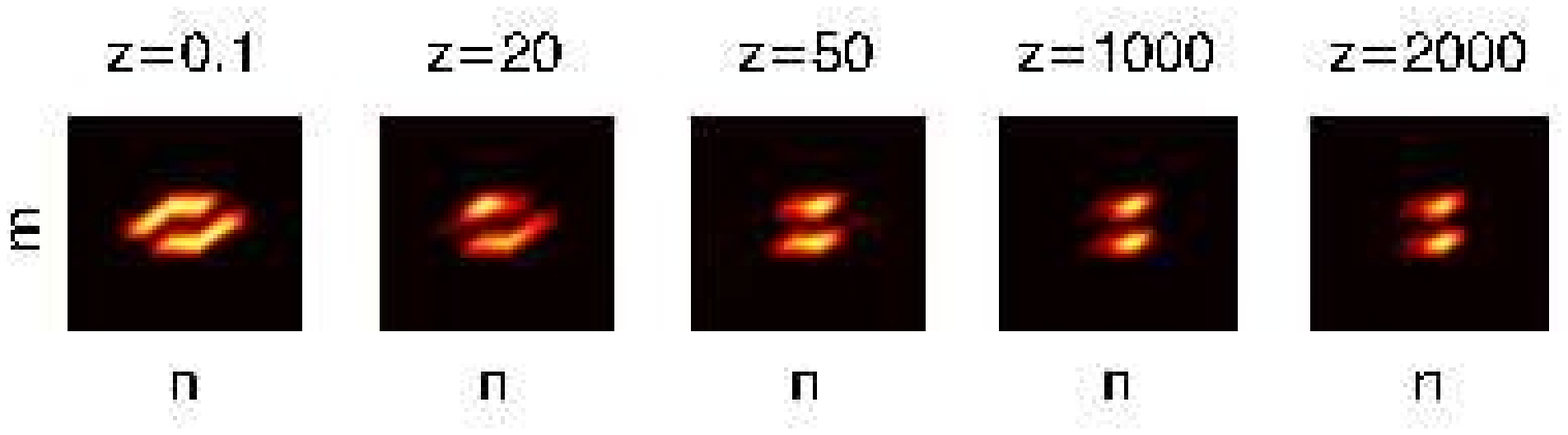}
\includegraphics[width=85mm]{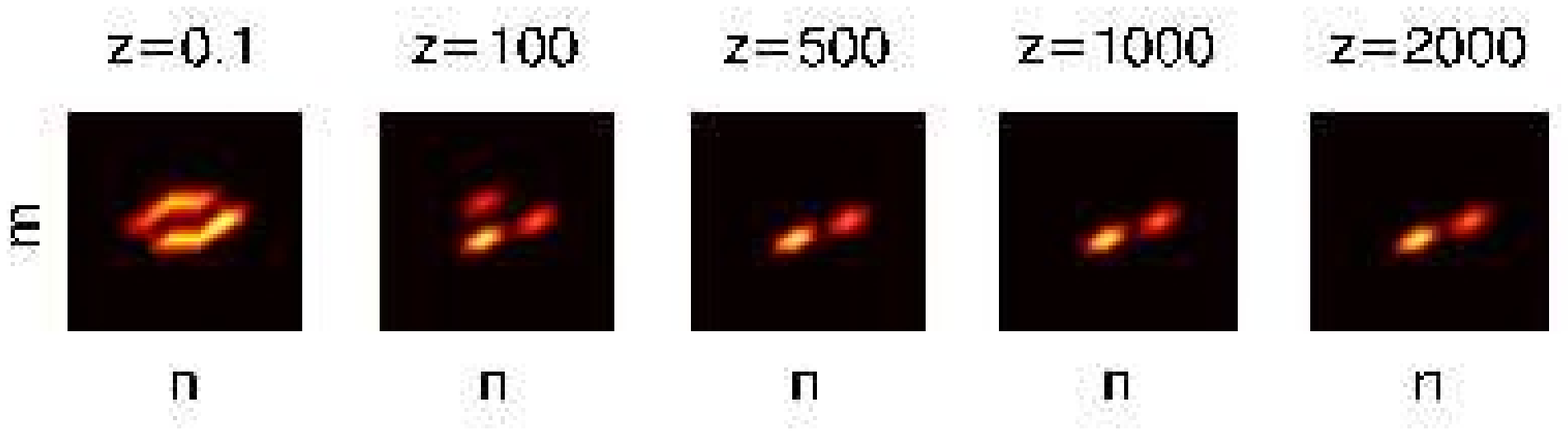}
\includegraphics[width=85mm]{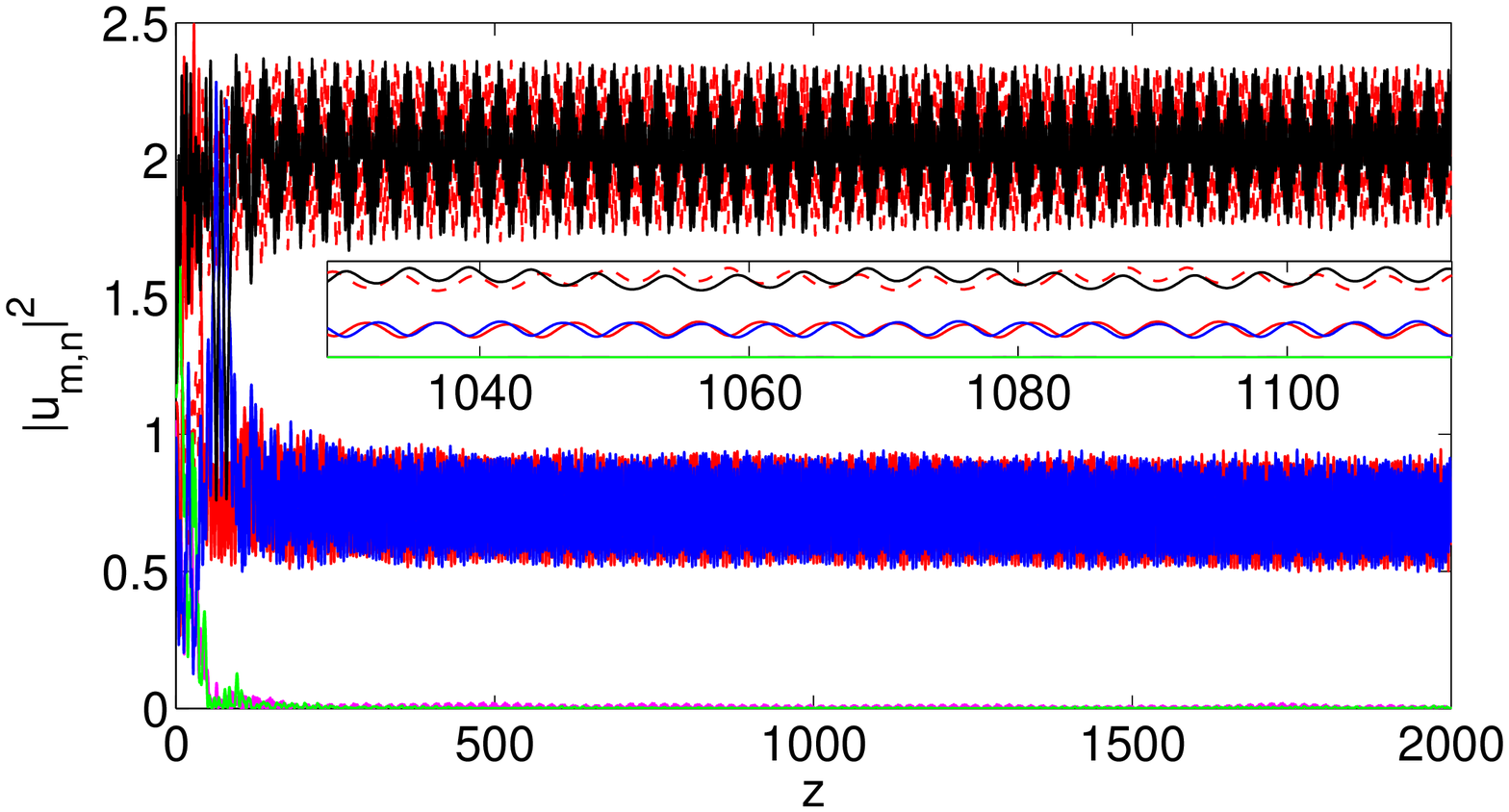}
\includegraphics[width=85mm]{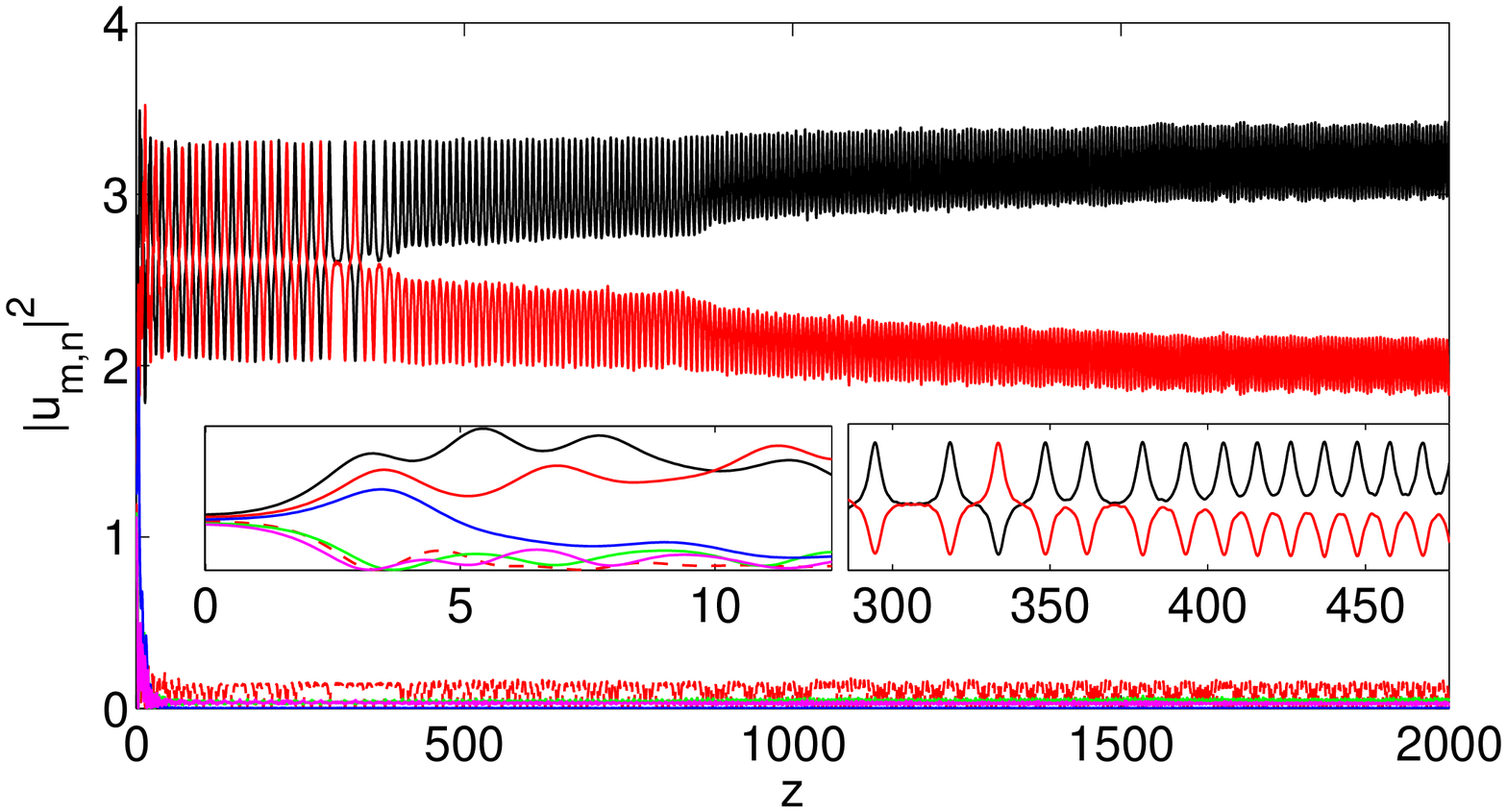}
\end{center}
\caption{(Color online) The evolution of the in-phase
six-site configuration with the honeycomb lattice
geometry from the top row of Fig. \ref{honey_6r}
is given above (left) for $\varepsilon=0.1$ and (right)
for $\varepsilon=0.3$.
As in the hexagonal case shown in Fig. \ref{hex_6ip_dyno}
for $\varepsilon=0.1$
a multi site structure persists over a long distance, although
now it is comprised of four sites, two pairs of out-of-phase
breathers with comparable amplitude (the phase structure is not shown,
but each pair is comparable with that of Fig. \ref{hex_6s2_dyno}).
This interesting difference inspired us to continue the dynamical evolution
for a longer distance, and the structure did indeed persist up to another
order of magnitude.
Even with the much larger perturbation of $25\%$
of the initial amplitude (not shown) as opposed to $5\%$,
a very similar four site structure persists for
a long distance, although the degeneration of the
other two sites is very rapid.
The same robustness to perturbation
is found
for $\varepsilon=0.3$, although a two site unequal amplitude
breather remains, which oscillates between in-phase and out-of-phase
(not shown, but same as the unequal amplitude breather in Fig. \ref{hex_6op_dyno}).}
\label{honey_6ip_dyno}
\end{figure*}

\begin{figure}[!ht]
\begin{center}
\includegraphics[width=85mm]{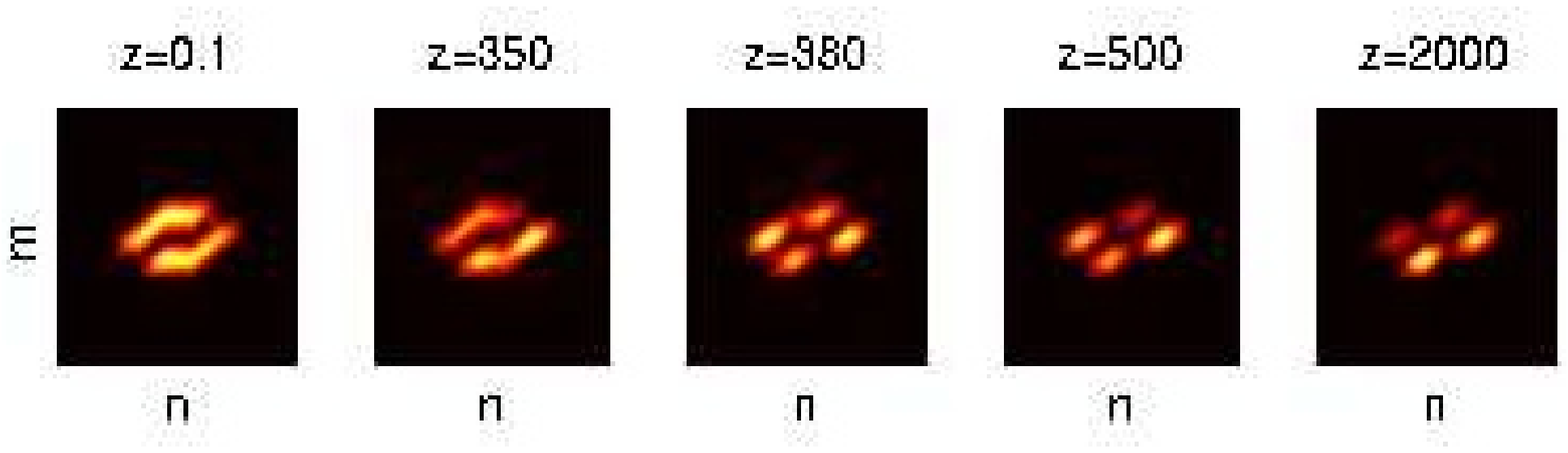}\\
\includegraphics[width=85mm]{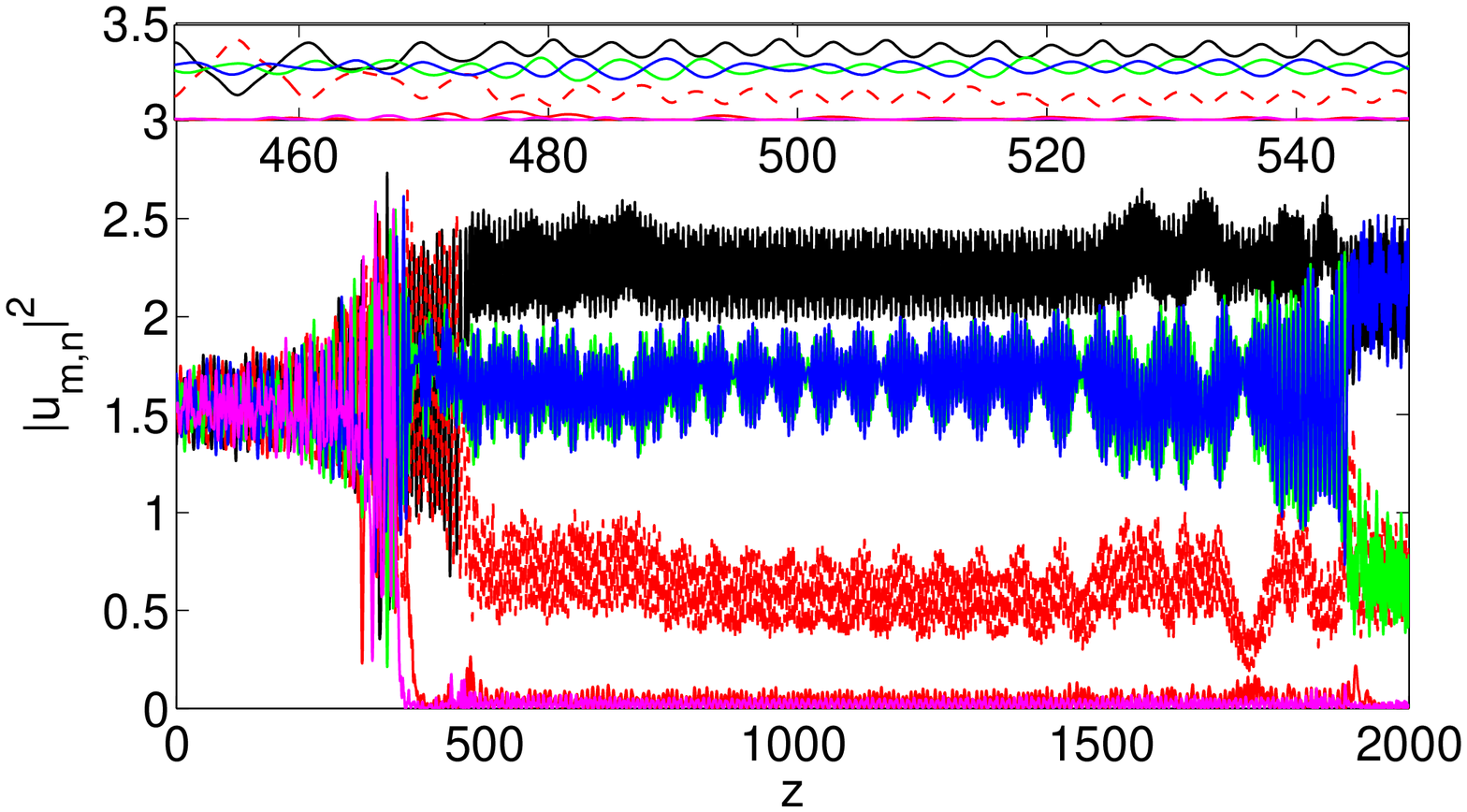}
\includegraphics[width=85mm]{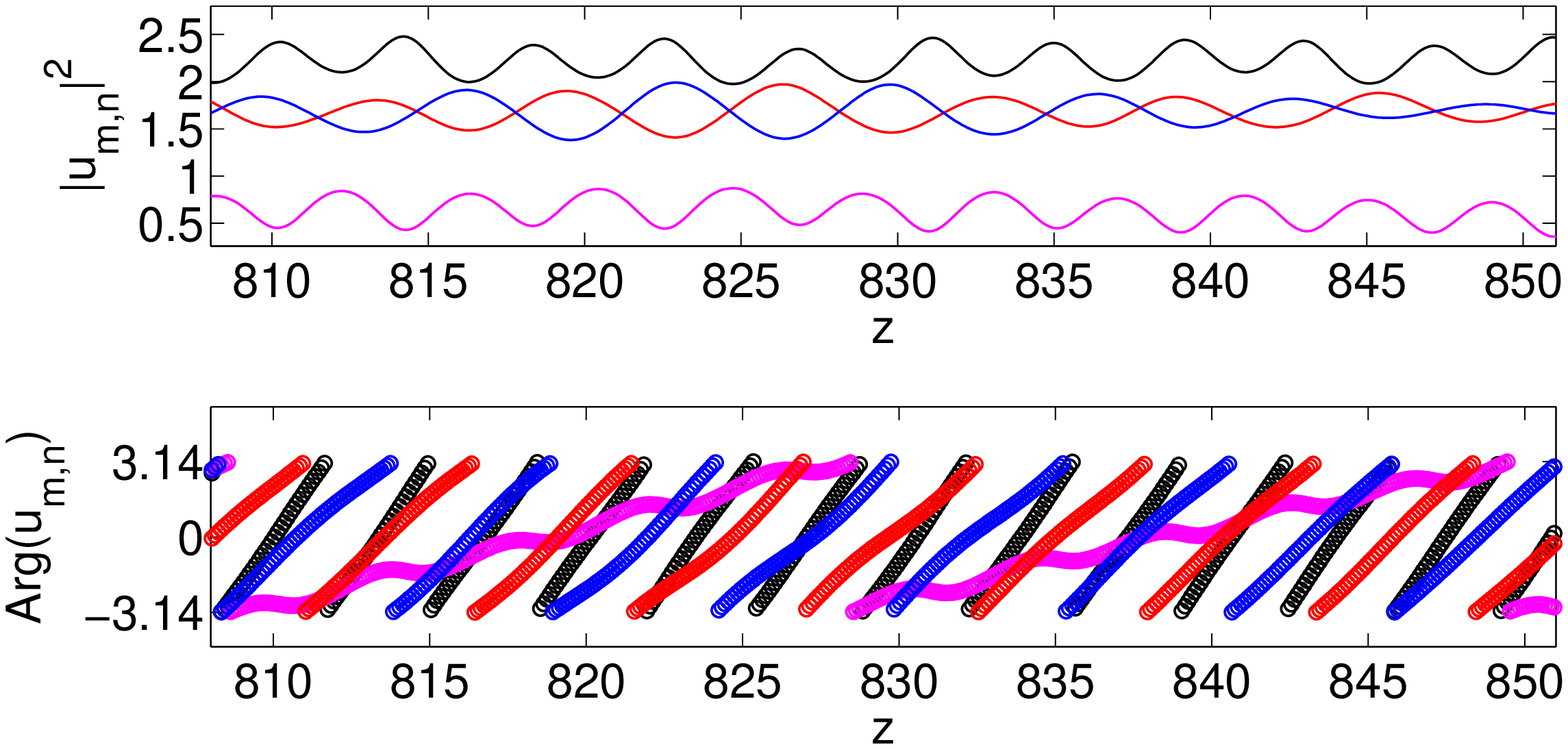}
\end{center}
\caption{(Color online) The six-site
doubly-charged honeycomb lattice
vortex for $\varepsilon=0.135$ from the bottom set
of panels in Figure \ref{honey_6s} is significantly
more stable than the
singly-charged counterpart. All original sites
remain populated for a long distance, up to $z=400$, and,
when the two sites eventually decrease in amplitude, the
remaining four reshape into a four-site breather.
Two of the sites remain close in amplitude and out-of-phase,
while one has larger and the other has smaller amplitude
and these oscillate between in-phase and out-of-phase in the
same manner as the others, such as those in Fig. \ref{hex_6op_dyno}
(see panels on the right).
The inset features a closeup image of the complex oscillations
of the four sites. At a very long distance, close to $z=2000$,
they reshape in amplitude and phase, becoming two pairs of
out-of-phase breathers,
like the in-phase hexapole from Fig. \ref{honey_6ip_dyno} ultimately does,
although the dynamics is not followed further
to see if this structure persists.}
\label{honey_6s2_dyno}
\end{figure}

\begin{figure*}[!ht]
\begin{center}
\includegraphics[width=85mm]{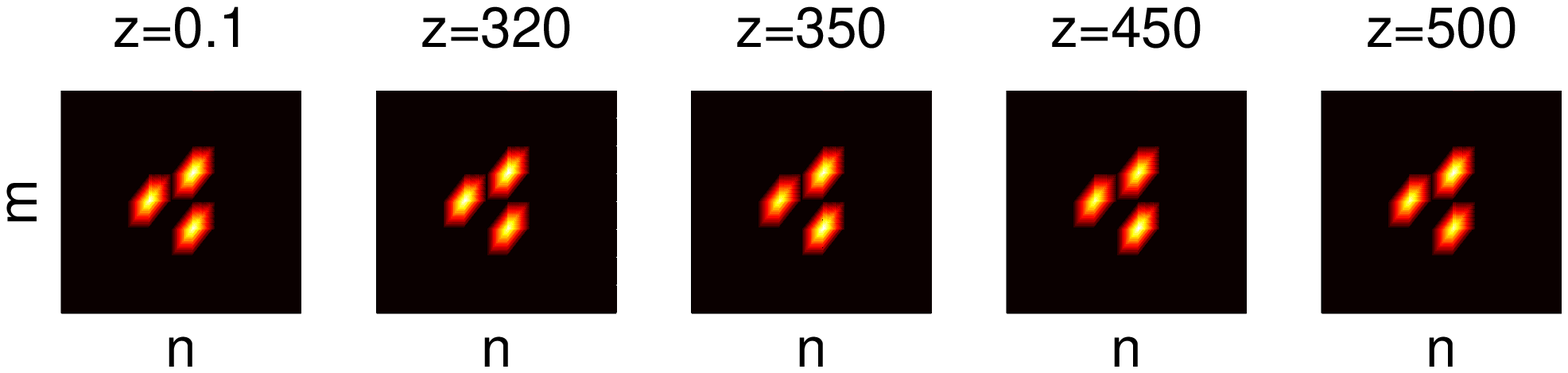}\\
\includegraphics[width=85mm]{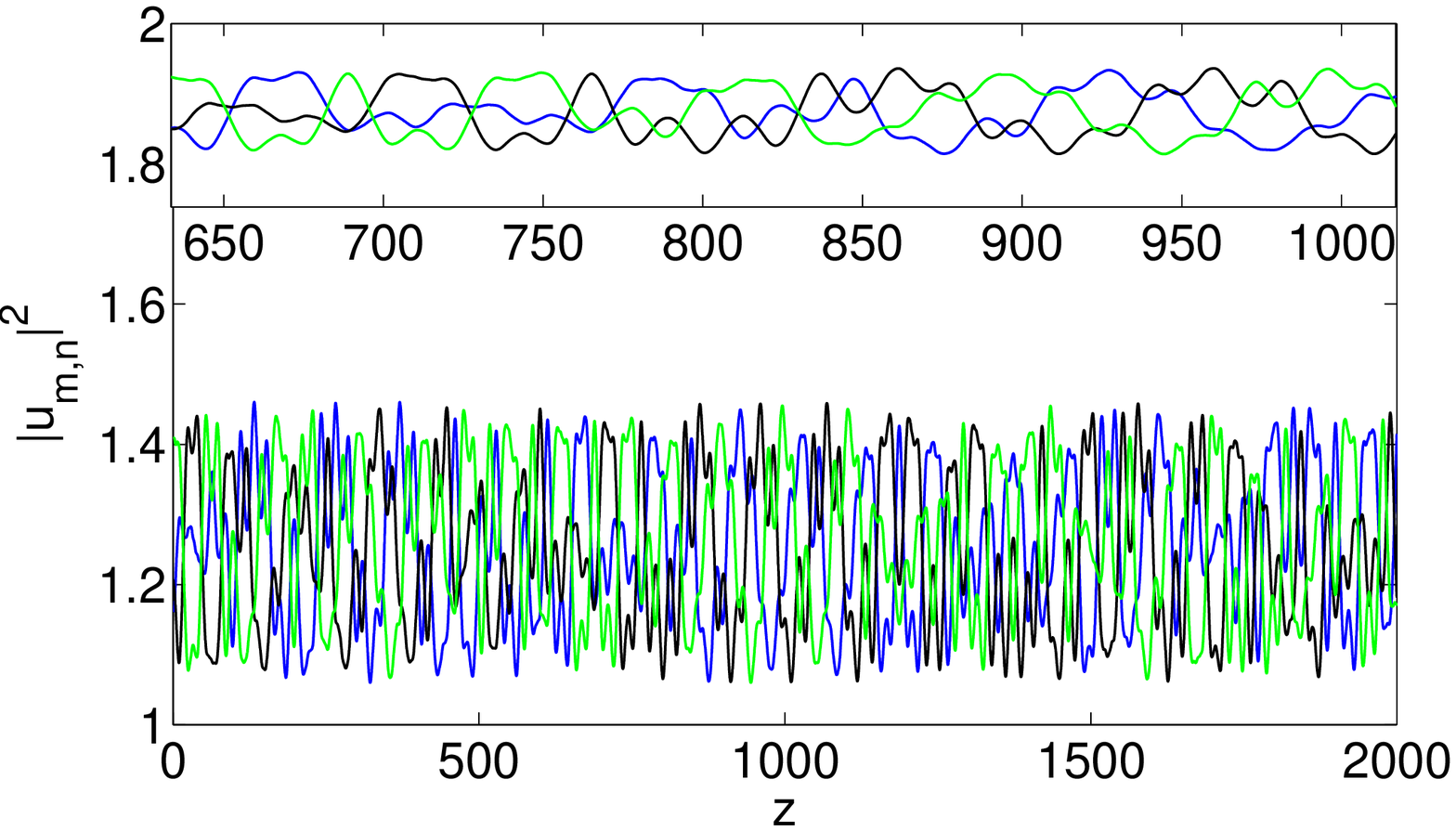}
\includegraphics[width=85mm]{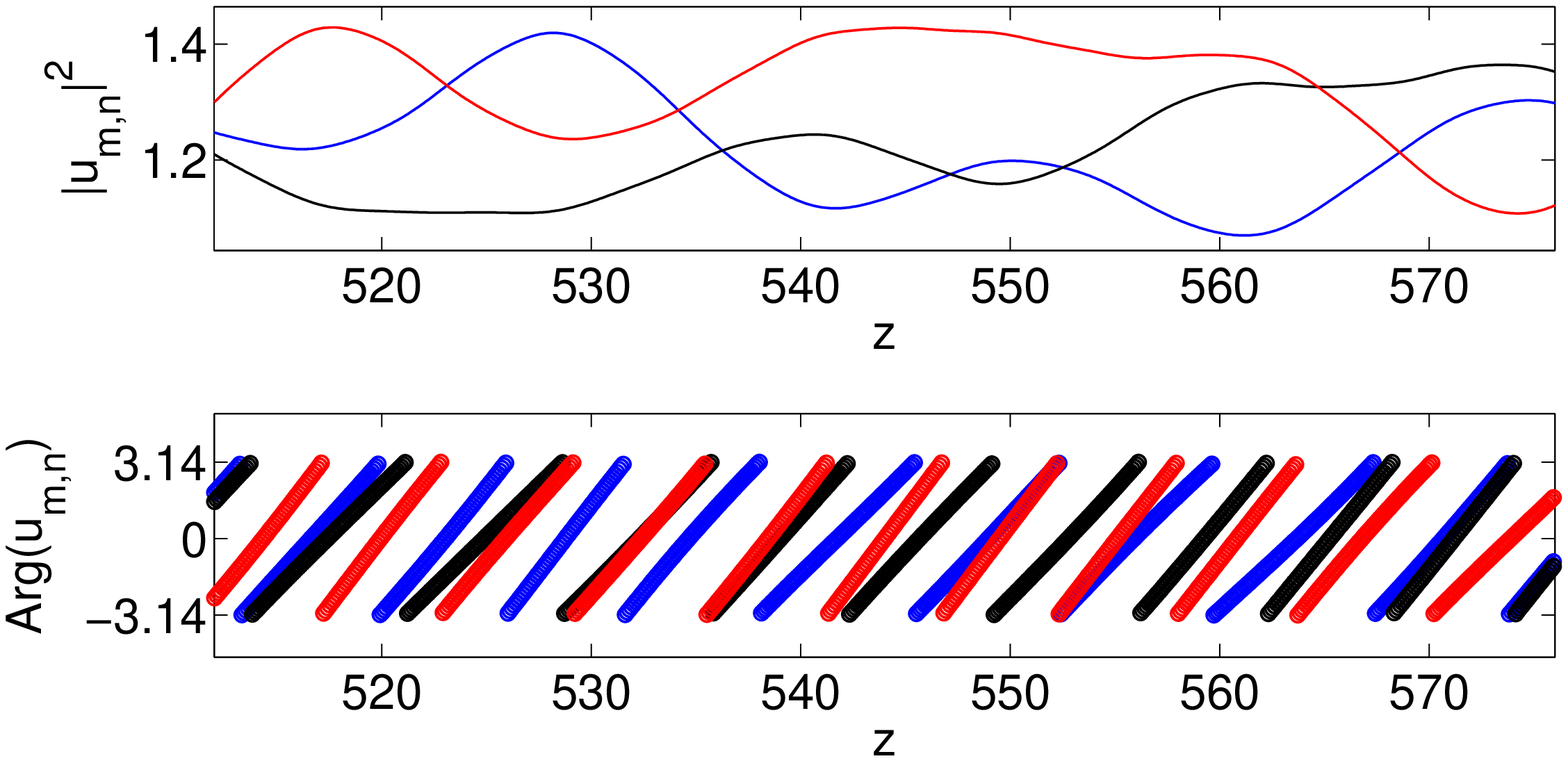}
\end{center}
\caption{(Color online) The dynamics for the $0,\pi,0$
honeycomb lattice configuration
 from the top rows of Fig. \ref{honey_3} for $\varepsilon=0.1$
This solution persists for very long propagation distances
despite the linear instability.  Moreover, the relative phase
structure persists, although the one site that is out-of-phase
with the other two oscillates from one to another among the
three (see the right panels).
}
\label{honey_3udu_dyno}
\end{figure*}


We now turn to the same configurations as above but in the
honeycomb geometry, as explored
in section \ref{hon_exstab}.
Interestingly, in this case,
for $\varepsilon=0.1$, and a $5\%$ perturbation, all configurations
result in multi-site breathing structures with up to four populated
sites for long propagation distances.
Since the dynamical evolution of the six-site configurations in the
hexagonal geometry all involve communications with the center site, it
is reasonable to hypothesize that this is a major contributor to the rather
significant differences observed below between
the nonlinear evolution presented
in this and the previous subsection.

First, we display the results of the evolution of a
real valued configuration from Fig. \ref{honey_6r} in
Fig. \ref{honey_6ip_dyno}.
The linearized system of the solution with
$\Delta \theta=0$ in Fig. \ref{honey_6ip_dyno} is strongly unstable with five
real pairs of eigenvalues, and the one with $\Delta \theta=\pi$
has all the same multiplied by the imaginary unity.
The dynamical evolution confirms the stability analysis and two sites decay
very rapidly for the in-phase configuration, while much more slowly for the
more stable out-of-phase one (not shown).
On the other hand, it is noteworthy that the four
sites persist for
long distances in each case and that the more linearly stable out-of-phase one
decays into three sites eventually, which have apparently uncorrelated
phases.  The resulting four site breather in the in-phase case for
$\varepsilon=0.1$ (left)
is actually comprised of two out-of-phase breather pairs, such as the one
in Fig. \ref{hex_6s2_dyno}, while for $\varepsilon=0.3$ (right)
the phases of the unequal amplitude breather pair oscillate between in-phase
and out-of-phase.
Also, as in the hexagonal
case of the $\Delta \theta=0$ family, we explored the sensitivity of the
nonlinear evolution to larger perturbation and coupling and found that
two sites robustly remain for the larger coupling $\varepsilon=0.3$, while four
remain for $\varepsilon=0.1$ (not shown). For this reason,
these solutions were continued
for an extra long propagation distance up to $z=2000$, and for
consistency and comparison the remaining cases in this setting
will also be continued for the same distances.


The instability of the discrete vortices from Fig. \ref{honey_6s}
results
in multiple sites persisting with
large amplitude oscillations for long
distances, ultimately
evolving to an out-of-phase
two-site breathing structure for the singly-charged ($\Delta \theta=\pi/3$)
one (not shown) and
a four-site structure for the doubly-charged ($\Delta \theta=2\pi/3$),
shown in  Fig. \ref{honey_6s2_dyno}.
This four site structure consists of an out-of-phase pair
close in amplitude and an unequal amplitude pair oscillating between
in-phase and out-of-phase (see right panels) until a very long
distance when they reshape
into two out-of-phase pairs (phase not shown). The latter part is
reminiscent of  the result of evolution of
the in-phase hexapole for small $\varepsilon$ given in
the left panels of Fig. \ref{honey_6ip_dyno}.

Finally, we
show the evolutions of the three-site configurations from
Fig. \ref{honey_3}. Figure \ref{honey_3udu_dyno}
displays the dynamics of an
unstable $\theta_i = 0,\pi,0$ solution.
The persistence for very long
distances of the three sites
for the smaller coupling prompted an investigation
of a solution with larger coupling of $\varepsilon=0.27$
from this family.
This turned out to
decay very rapidly to a single site (not shown).
In the smaller coupling case, an intricate breathing pattern emerges
which apparently
converts the mode into a three-site breather
(rather than a three-site stationary solution).
The three-site configuration with $\Delta \theta = 0$ is not shown,
but again here all three sites survive for a long propagation
distance for $\varepsilon=0.1$.
This does not necessarily
contradict the linear instability, since the configuration deviates almost
immediately in terms of amplitude distribution. There is no clear
correlation in the phases in this case. Again the persistence of all
three sites for $\varepsilon=0.1$ prompted investigation for a larger coupling
$\varepsilon=0.3$ and again a single site ultimately remained, although
in this case two sites also persisted for
a significant distance
before
the ultimate degeneration into a single-site waveform.
For the last three site configuration
the same consideration of the coupling
arises, since this configuration is unique among those considered
here, in the sense that a considerably larger coupling strength is required
for the imaginary eigenvalues to collide with the continuous spectrum
and the instability of this state to occur.
Even with the very mild
instability
when the first imaginary pair collides with the phonon band at the very large
coupling value of $\varepsilon=0.43$, the dynamics
clearly illustrate
the oscillatory instability.
The original configuration
persists until $z=30$, ultimately concentrating
primarly on a single site for long propagation distances.
Aside from the six-site in-phase configuration, this is the only one for
which the dynamics are qualitatively similar in the honeycomb and
hexagonal geometries.
Breathers remained for certain parameter values for all other
configurations considered.
The relative phases of the two-site breathers
which
recurred in many of the simulations suggest that these may
exist as potentially stable time-periodic solutions.

\section{Conclusions}

In the present work we have examined both discrete soliton and discrete
vortex configurations on non-square lattices for a prototypical
discrete nonlinear Hamiltonian evolution equation (namely, the DNLS equation)
of diverse interest to various areas of nonlinear optics and potentially of
atomic
physics. We studied, in particular, three- and six-site configurations in
non-square lattice geometries in the case that each node has six neighbors
(hexagonal lattice), as well as in the case that each node has three neighbors
(honeycomb lattice). Theoretical predictions for the stability of six-site real
configurations were the same in each case and the in-phase version was
strongly unstable (and generally all configurations with any pair of
in-phase nearest neighbors will be unstable), while the out-of-phase
configuration was subject only to weak oscillatory instabilities stemming
from complex quartets of eigenvalues that arise when imaginary eigenvalues
with negative
Krein signature collide with the continuous spectrum (of positive
Krein signature).
On the other hand, among complex solutions, we highlighted the
cases of topological, singly- and doubly-charged
structures (discrete vortices).
In both the hexagonal and honeycomb geometries it was found
that the former configuration is strongly unstable, while the
latter may be stable for sufficiently weak couplings.
It is also relevant in this context to point out that our
results
were presented for the case of a focusing nonlinearity,
but it is
straightforward to extend them to the defocusing case.
There, it is expected that these features will be reversed, i.e., the
vortex of charge $S=1$ will be stable, while that of $S=2$
will be unstable in contrast to the effect of such a transformation
on the fundamental four-site contour in a square lattice.
Similarly the in-phase structure will be
stabilized, while the out-of-phase one will be destabilized
which, however, is consistent with
the case of the square lattice.
This can be inferred by a staggering transformation along the
one-dimensional contour of excited sites, which changes by $\pi$
the phase of, say, just the odd (or equivalently just the even)
sites of the contour.
For the three-site configurations, on the other hand, there was
a difference between the honeycomb and hexagonal geometries which
arises due to the fact that such a configuration
occurs in the honeycomb case
comprised of next nearest neighbors and therefore
the eigenvalues grow at a higher order in $\varepsilon$: in fact, linearly
(compared to a growth proportional to $\sqrt{\varepsilon}$ in the hexagonal
case). Nevertheless, in all
cases, our theoretical predictions were confirmed qualitatively,
but also quantitatively (at least for small $\varepsilon$) by numerical
continuation/bifurcation results.

Finally, the dynamics of these states revealed a number of
somewhat unexpected features. Unlike the expectation of square geometries
that the relevant configurations may typically degenerate to
a single-site solitary excitation, we have observed
in various cases, that not only
can the configuration reduce
to waveforms with a larger number of excited sites, but also it
can even preserve its original amplitude profile in terms of
the number of excited sites -- exhibiting
breathing oscillations in the amplitude of these sites,
or an internal reshaping of their relative phase.
A breather composed of two-sites having comparable amplitudes
and being usually out-of-phase with one another
recurs in the dynamics of several
solutions, as well as one with two uneven amplitudes in which the
phases oscillate between in-phase and out-of-phase according to
whether the amplitudes are closer or further apart, respectively.
Also, in some
cases there are very particular amplitude structures with
no apparent phase correlation, such as
the three-site structure
that persist over long
distances after the decomposition of the six-site
in-phase configuration in the hexagonal geometry.
It is also worthwhile to highlight the sensitive
dependence of the evolution on the coupling parameter. For large values
of the coupling, the degeneration to a single-site excitation apparently
becomes more likely due to the fact that many of the multi-site configurations
disappear through
bifurcations, as has been discussed, e.g.,
in one-dimensional settings in \cite{alfimov}.

There is a number of interesting directions suggested by
these results for future studies.
It would be relevant
to examine, along lines similar to the earlier work of \cite{kouk},
whether the conclusions presented herein are also qualitatively
similar to what can be found for discrete solitons and vortices
in two-dimensional Klein-Gordon chains.
It would also be relevant to
consider the continuum analogs of these results, either for cubic
or for saturable
nonlinearities (appearing, e.g., in photorefractive crystals)
and see how the latter compare to the discrete theory, especially as concerns
the qualitative structure of the linearization spectrum and the solutions
predicted to be potentially stable.
Another direction would be to explore in more
detail the nonlinear dynamics,
to determine the exact mechanism which supports the complex
breathing dynamical structures, identify exact breather
solutions, and to explain the sensitivity of the
evolution on the coupling parameter.
Finally,
another interesting extension would be to consider such non-square
lattices in a fully three-dimensional setting, including
hexagonal-close-packed configurations.
Studies along some of these directions are presently underway and
will be reported in future publications.

\begin{acknowledgments}
PGK gratefully acknowledges support from NSF-DMS-0505663, NSF-DMS-0806762
and NSF-CAREER, as well as from the Alexander von Humboldt Foundation. 
The work of IK was supported by a UK EPSRC Science and Innovation Award.
\end{acknowledgments}

\end{document}